\documentclass[11pt, letterpaper,openany]{article}
\pdfoutput=1
\usepackage[table]{xcolor}
\usepackage{amsmath}
\usepackage{amssymb}
\usepackage{amsthm}
\usepackage{graphicx}
\usepackage{multirow}
\usepackage{booktabs}
\usepackage{tabularx}
\usepackage{hyperref}
\usepackage{url}
\usepackage{algorithm}
\usepackage{algorithmic}
\usepackage{caption}
\usepackage{subcaption}
\usepackage{booktabs}
\usepackage{authblk}
\usepackage{appendix}
\usepackage[utf8]{inputenc}
\usepackage[T1]{fontenc}
\usepackage[english]{babel}
\usepackage{geometry}
\usepackage{xcolor}
\usepackage{fancyhdr}
\usepackage{microtype}
\usepackage{parskip}
\usepackage{tabularx}
\usepackage{enumitem}
\usepackage{float}
\usepackage[numbers,sort]{natbib}
\hypersetup{
    colorlinks=true,
    linkcolor=blue,
    filecolor=magenta,      
    urlcolor=cyan,
    citecolor=red,
}

\definecolor{headercolor}{RGB}{0, 50, 100}
\title{Potential of Multimodal Large Language Models for Data Mining of Medical Images and Free-text Reports }

\date{}
  
\newcommand*\samethanks[1][\value{footnote}]{\footnotemark[#1]}
\author[1]{Yutong Zhang \thanks{Co-first authors.}}
\author[2,3]{Yi Pan \samethanks}
\author[4]{Tianyang Zhong \samethanks}

\author[4]{Peixin Dong \thanks{Co-second authors.}}
\author[5]{Kangni Xie \samethanks}
\author[6,7]{Yuxiao Liu \samethanks}

\author[2]{Hanqi Jiang}
\author[2]{Zhengliang Liu}

\author[4]{Shijie Zhao}
\author[4]{Tuo Zhang}
\author[3]{Xi Jiang}
\author[6,8,9]{Dinggang Shen}
\author[2]{Tianming Liu}
\author[1]{Xin Zhang\thanks{Corresponding author. E-mail: xzhang@nwpu.edu.cn}}

\affil[1]{Institute of Medical Research, Northwestern Polytechnical University, Xi'an 710072, China}
\affil[2]{School of Computing, University of Georgia, GA, USA}
\affil[3]{School of Life Science and Technology, University of Electronic Science and Technology of China, Chengdu 611731, China}
\affil[4]{School of Automation, Northwestern Polytechnical University, Xi'an 710072, China}
\affil[5]{School of Electronics and Information, Northwestern Polytechnical University, Xi'an 710072, China}
\affil[6]{School of Biomedical Engineering, ShanghaiTech University, Shanghai 201210, China}
\affil[7]{Lingang Laboratory, Shanghai, 200031, China}	
\affil[8]{Shanghai United Imaging Intelligence Co., Ltd.}
\affil[9]{Shanghai Clinical Research and Trial Center}

\begin{document}

\maketitle

% \newpage

\begin{abstract}
% XXX(ZYT)

Medical images and radiology reports are essential for physicians to diagnose medical conditions, emphasizing the need of quantitative analysis for clinical decision-making. However, the vast diversity and cross-source heterogeneity inherent in these data have posed significant challenges to the generalizability of current data-mining methods. Recently, multimodal large language models (MLLMs) have revolutionized numerous domains, significantly impacting the medical field. Notably, Gemini-Vision-series (Gemini) and GPT-4-series (GPT-4) models have epitomized a paradigm shift in Artificial General Intelligence (AGI) for computer vision, showcasing their potential in the biomedical domain. In this study, we evaluated the performance of the Gemini, GPT-4, and 4 popular large models for an exhaustive evaluation across 14 medical imaging datasets, including 5 medical imaging categories (dermatology, radiology, dentistry, ophthalmology, and endoscopy), and 3 radiology report datasets. The investigated tasks encompass disease classification, lesion segmentation, anatomical localization, disease diagnosis, report generation, and lesion detection. Our experimental results demonstrated that Gemini-series models excelled in report generation and lesion detection but faces challenges in disease classification and anatomical localization. Conversely, GPT-series models exhibited proficiency in lesion segmentation and anatomical localization but encountered difficulties in disease diagnosis and lesion detection. Additionally, both the Gemini series and GPT series contain models that have demonstrated commendable generation efficiency. While both models hold promise in reducing physician workload, alleviating pressure on limited healthcare resources, and fostering collaboration between clinical practitioners and artificial intelligence technologies, substantial enhancements and comprehensive validations remain imperative before clinical deployment.

\end{abstract}

% \tableofcontents

% \newpage

% \listoffigures

% \newpage

\section{Introduction}

% XXX(PY)

% In the advancements within the domain of natural language processing (NLP), the progression from monomodal (i.e., text only) to multimodal large language models (MLLMs) signifies a paradigm shift in artificial general intelligence (AGI) research. Notably, models such as ChatGPT and its derivatives have laid the groundwork by demonstrating exceptional capabilities in generating and interpreting text. Yet, the exploration of these models’ competencies in integrating and interpreting visual data, especially in domains requiring a high degree of specialization like the biomedical and medicine fields, heralds a new frontier in AI application. Among the forerunners of this innovative thrust are GPT-4-series and Gemini-Vision-series models (GPT-4 and Gemini for short), two state-of-the-art series of multimodal models that epitomize the fusion of linguistic and visual information processing.
Recent advancements in natural language processing (NLP) have shifted from monomodal (i.e., text-only) to multimodal large language models (MLLMs), marking a significant paradigm shift in artificial general intelligence (AGI) research. To begin with text, language has always been a distinguishing feature of human intelligence from animals'. With the advancement of artificial intelligence (AI), particularly in the field of NLP, machines are becoming increasingly capable of understanding and processing language. Over the past few years, pre-trained language models (PLMs) based on self-attention mechanisms and the Transformer framework~\cite{vaswani2017attention} have emerged and rapidly gained popularity. PLMs can learn general language representations from large-scale data in an unsupervised manner, which facilitates various downstream NLP tasks without the need for retraining new models~\cite{han2021pre}. It is noteworthy that when the scale of training data and parameters exceeds a certain threshold, language models exhibit significant performance improvements and acquire capabilities absent in smaller models, such as context learning. We refer to such models as large language models (LLMs). Large language models (LLMs) like GPT-3~\cite{NEURIPS2020_1457c0d6} and its derivatives (e.g., InstructGPT~\cite{NEURIPS2022_b1efde53}, also known as ChatGPT), Llama series (i.e., Llama~\cite{touvron2023llama}, Llama 2 \cite{touvron2023llama2}, and Llama 3~\cite{llama3modelcard}), and PaLM series (i.e., PaLM \cite{chowdhery2023palm}) and PaLM 2 \cite{anil2023palm}), have laid the groundwork by demonstrating exceptional text interpretation and generation capabilities~\cite{zhao2023survey}. 

The advent of the era of LLMs has brought us closer to the dawn of AGI~\cite{LIU2023100017,NEURIPS2023_1190733f}. Academia and industry are experiencing vibrant competition and diverse developments, from the early Google T5 model~\cite{JMLR:v21:20-074} to the highly acclaimed OpenAI's GPT series today. The parameter scales of these models have long surpassed the billion-level mark, and their generative and learning capabilities of LLMs are revealing emergent abilities~\cite{wei2022emergent} and increasing being applied across various sectors. LLMs have demonstrated exceptional proficiency in understanding and generating natural language, providing foundational solutions for specific domains such as law~\cite{10.1145/3630106.3659048,cui2023chatlaw,izzidien2024llm}, education~\cite{shu2024llms,10.1007/978-3-031-36336-8_4,Irfan2023}, and public healthcare~\cite{zhang2024generalizable,liu2023evaluating,10433180}. However, each domain deals with unique challenges, and directly using pre-trained LLMs may not yield ideal results. Fine-tuning these models, considering their inherent complexity, enables them to better adapt to downstream tasks, which is a key approach to leveraging large models \cite{zhong2023chatradio,zhao2023ophtha,yang2023harnessing}.

Despite their exceptional proficiency in zero/few-shot reasoning in most NLP tasks, LLMs face challenges in processing visual information as they can only understand discrete text. Meanwhile, large-scale visual foundation models have made significant advancements in perception, leading to the gradual integration of monomodal LLMs and visual models, ultimately giving rise to the emergence of MLLMs~\cite{WANG2023100047}. MLLMs are models based on LLMs that can receive and reason with multimodal information, extending beyond the traditional single "language modality" to include "image," "speech," and other "multimodal" data. Among these, Gemini \cite{anil2023gemini} and GPT-4 \cite{achiam2023gpt} are notable examples. Gemini combines language and visual information processing, while GPT-4 enhances the understanding and generation of visual data, garnering widespread attention. From the perspective of developing AGI, MLLMs may represent a step forward compared to LLMs, as they align more closely with human ways of perceiving the world, are more user-friendly, and generally support a broader range of tasks \cite{yin2023survey}.

However, the exploration of these models’ abilities to integrate and interpret visual data, particularly in highly specialized domains such as biomedicine, signifies a new frontier in AI application. Notable among these advancements are the GPT-4-series and Gemini-Vision-series models (namely GPT-4 and Gemini for the rest of this article), which epitomize the fusion of linguistic and visual information processing.

% This research conducts a meticulous comparative analysis of GPT-4 and Gemini, focusing on their application in biomedical image analysis. GPT-4, an advanced extension of the monomodal ChatGPT model, and Gemini, a similarly advanced multimodal model, are both designed to comprehend and analyze information across textual and visual dimensions. This analysis seeks to unravel the extent to which these multimodal large models can navigate the complexities of visual data within the biomedical domain, an exploration that could considerably widen their applicability and effectiveness.

This research conducts a meticulous comparative analysis of GPT-4 and Gemini, focusing on their application in biomedical image analysis. And we also design experiments on the popular models including Yi, Claude, and Llama 3, to evaluate these LLMs' textual comprehension and MLLMs' multimodal comprehension compared to GPT-4 and Gemini. GPT-4, an advanced extension of the monomodal ChatGPT model from OpenAI, and Gemini, a similarly advanced multimodal model from Google DeepMind, are designed to comprehend and analyze information across textual and visual dimensions. This study explores how effectively these multimodal models can handle the complexities of visual data within the biomedical domain, potentially broadening their applicability and effectiveness.

% The evaluation methodology employed in this study encompasses a series of rigorous tests designed to assess the models' accuracy, efficiency, and adaptability in interpreting and leveraging visual information for biomedical purposes. By scrutinizing the performance of GPT-4 and Gemini in tasks such as medical image classification, anomaly detection, and data synthesis, this paper aims to illuminate the strengths and limitations of each model. Additionally, it endeavors to provide insights into how these MLLMs could be further optimized for specialized applications.

The evaluation methodology includes a series of rigorous tests to assess the models' accuracy, efficiency, and adaptability in interpreting and leveraging visual information for biomedical purposes. By examining the performance of GPT-4 and Gemini in tasks such as medical image classification, anomaly detection, and data synthesis, this paper highlights each model's strengths and limitations. Additionally, it offers insights into optimizing these MLLMs for specialized applications.

% In doing so, this investigation not only highlights the groundbreaking potential of integrating advanced AI models like GPT-4 and Gemini into the realm of biomedical analysis but also sets a benchmark for future research in the field. By comparing these models, the study contributes valuable knowledge to the ongoing discourse on enhancing AI's multimodal capabilities, particularly in sectors where the amalgamation of text and visual data is critical. This research underscores the transformative impact that such advancements could have on medical diagnostics, treatment planning, and the broader biomedical field, marking a significant step toward the realization of fully integrated AGI systems in specialized domains.

This investigation not only showcases the groundbreaking potential of integrating advanced AI models like GPT-4 and Gemini into biomedical analysis but also sets a benchmark for future research in the field. By comparing these models, the study provides valuable knowledge to the ongoing discourse on enhancing AI's multimodal capabilities, especially in sectors where combining text and visual data is crucial. This research underscores the transformative impact these advancements could have on medical diagnostics, treatment planning, and the broader biomedical field, marking a significant step toward realizing fully integrated AGI systems in specialized domains. Overall, the main contributions of our work are summarized as follows:

\begin{enumerate}
    \item We provide a detailed comparative analysis of GPT-4 and Gemini models, specifically focusing on their application in biomedical image analysis, highlighting their strengths and limitations across multiple tasks such as disease classification, lesion segmentation, and report generation.
    \item Our study employs rigorous evaluation methodologies to assess the accuracy, efficiency, and adaptability of these models in interpreting and leveraging visual information, offering insights into their potential optimization for specialized biomedical applications.
    \item By integrating advanced AI models into biomedical analysis, our research underscores their transformative impact on medical diagnostics, treatment planning, and the broader biomedical field, setting a benchmark for future AGI system developments in specialized domains.
\end{enumerate}

% \begin{figure}[!h]
%     \centering    \includegraphics[width=1\textwidth,height=!,keepaspectratio]{figures/main-v3.pdf}
%     \caption{\textbf{Schematic Overview of the Evaluation Methodology.} We crafted effective prompts to feed diverse data categories into a large language model, subsequently conducting manual evaluations and engineering metric calculations for each set of results.}
%     \label{fig:main}
% \end{figure}

\section{Related work}
\label{sec:related}

\subsection{Large Language Models}
\label{subsec:Large Language Models}

% XXX(LYX)
% Briefly why LLM emerges

With the increasing GPU computing ability and  training data size, a series of Transformer \cite{vaswani2017attention} based pre-trained LLMs have emerged. Pre-trained LLMs can be grouped into encoder based \cite{devlin2018bert,lan2019albert,lewis2019bart,liu2019roberta}, decoder based \cite{du2021glm,brown2020language,zhang2022opt,touvron2023llama}, and encoder-decoder based models \cite{tay2022ul2,chung2022scaling,raffel2020exploring,lewis2019bart}.
Encoder based LLMs are better at analyzing and classifying text content, including semantic feature extraction and named entity recognition. The first encoder based pre-trained LLM is the Bidirectional Encoder Representations from Transformers (BERT) \cite{devlin2018bert}. BERT uses bidirectional language encoders with specially designed context-aware and mask prediction tasks on large-scale unlabeled text data. 
Following BERT \cite{devlin2018bert}, roBERT \cite{liu2019roberta} further improves the performance by updating its training method, such as expanding the batch size, training on larger data, and eliminating BERT's next-sentence prediction task training method.

Although encoder based LLMs can effectively extract sentence features,  they can not perform well on zero-shot or few-shot tasks, which are vital for LLMs. Generative Pre-trained Transformer (GPT) in the other view, uses the auto-regressive task for training and acquires better generalization ability in both zero-shot and few-shot tasks \cite{wang2022language}. Decoder based models use the similar multi-head self-attention mechanism to the one in the encoder, but the attention mask prevents the model from attending to future positions, ensuring that the predictions for position $i$ can only depend on the known outputs at positions less than $i$. Radford et al. find the scaling (model size, dataset size or both) can largely improve the decoder-based LLMs capacity \cite{radford2019language}. This makes decoder framework widely used in scaling LLMs (ChatGPT as the most typical example). Although the scaling LLMs are based on similar deep learning architecture and training algorithm compared with smaller LLMs, they exhibit new emergent capabilities that do not appear before. For example, GPT-3 \cite{brown2020language} can solve few-shot tasks through in-context learning, whereas GPT-2 cannot do well. Another remarkable application of scaling LLMs is ChatGPT that adapts the LLMs  for dialogue, which presents an amazing conversation ability with humans and can solve different complex tasks by $instruction-tuning$ \cite{ouyang2022training} and $chain-of-thought$ (CoT) \cite{wei2022chain}. With instruction tuning, LLMs are enabled to follow the task instructions for new tasks without using explicit examples, thus having an improved zero-shot ability which is vital for solving different tasks. CoT strategy can ease the model solving the difficult task by dividing it into multiple reasoning steps, then, LLMs can solve such tasks by utilizing the prompting mechanism that involves intermediate reasoning steps for deriving the final answer. 

There are also works \cite{tay2022ul2,chung2022scaling,raffel2020exploring,lewis2019bart} trying to build encoder-decoder based LLMs to take full advantage of encoder and decoder based LLMs.  Encoder-decoder based LLMs are mainly used to handle tasks that require precise mapping between input and output, such as machine translation, text summarization, etc. In these tasks, it is very important to understand the precise content of the input and generate specific output accordingly. Models trained based on this architecture can generally only be applied to certain specific tasks. For example, an encoder-decoder LLM specially trained for machine translation may not be suitable for direct use in text summarization or other types of tasks. This makes encoder-decoder based LLMs mostly used in specific field, instead of the general domain like decoder based LLMs.

To summarize, with scaling model and dataset, the zero-shot and few-shot become key capabilities for LLMs. The decoder based LLMs have greater flexibility compared with encoder and ecoder-decoder based LLMs. Models trained based on decoder only architecture can handle many different types of text generation tasks, such as question and answer, translation, etc., without the need for special training or adjustment for each task, thus making it more general in the practical.

\subsection{Multimodal Large Language Models}
\label{subsec:multi-modal-models}

% XXX(JHQ)

In recent years, LLMs have achieved significant progress \cite{zhao2023survey}. By scaling up model and data sizes, LLMs have demonstrated extraordinary emergent abilities, such as instruction following \cite{peng2023instruction}, in-context learning \cite{brown2020language}, and chain-of-thought reasoning \cite{wei2022chain}. Concurrently, Large Vision Models (LVMs) have also made substantial advancements \cite{kirillov2023segment, shen2024aligning, zhang2022dino, oquab2023dinov2}. MLLMs, as a natural extension, leverage the complementary strengths of LLMs and LVMs \cite{yin2023survey}.

MLLMs typically consist of a modality encoder, a pre-trained LLM, and a modality interface that bridges the gap between different modalities \cite{yin2023survey}. The modality encoder processes inputs from various modalities, such as images, videos, and audio, transforming them into representations that the LLM can comprehend. The pre-trained LLM is responsible for understanding and reasoning over these representations. The modality interface serves as a bridge, aligning and fusing information from different modalities into the LLM.

The training process of MLLMs generally involves three stages: pre-training to align modalities and learn multimodal knowledge, instruction tuning to enable generalization to new tasks, and alignment tuning to adapt the model to specific human preferences \cite{yin2023survey}. During the pre-training stage, researchers utilize large-scale image-text paired datasets, such as LAION \cite{schuhmann2022laion} and CC \cite{sharma2018conceptual, changpinyo2021conceptual}, to train the model to learn the alignment between different modalities and acquire rich world knowledge. In the instruction tuning stage, the model is fine-tuned on instruction datasets, such as LLaVA-Instruct \cite{liu2023visual} and ALLaVA \cite{chen2024allava}, to enhance its generalization ability to new tasks. During the alignment tuning stage, the model is trained using human feedback data, such as LLaVA-RLHF \cite{sun2023aligning} and RLHF-V \cite{yu2023rlhf}, to better adapt to human preferences and generate more accurate outputs with fewer hallucinations.

In addition to these open-source MLLMs, several commercial companies have introduced powerful closed-source models, such as OpenAI's GPT-4 \cite{openai2023gpt4}, Anthropic's Claude-3 \cite{claude2023}, and Microsoft's KOSMOS-1 \cite{peng2023kosmos1}. These models have demonstrated remarkable capabilities in handling multimodal tasks, such as generating stories based on images and performing mathematical reasoning without OCR. Their emergence has greatly promoted the development of MLLMs and provided new ideas and directions for researchers in the field.

To evaluate and compare the performance of different MLLMs, researchers have developed various benchmarks and methods. For closed-set problems (i.e., limited answer options), benchmarks such as MME \cite{fu2023mme}, MMBench \cite{liu2023mmbench} and MM-VET \cite{yu2023mm} provide comprehensive and fine-grained quantitative comparisons. For open-set problems (i.e., flexible and diverse answer options), evaluation methods such as human scoring \cite{zhu2023minigpt}, GPT scoring \cite{yang2023dawn} and case studies \cite{wen2023road} offer qualitative analyses of MLLMs' generative capabilities from different perspectives.

\subsection{Medical MLLMs}
\label{subsec:medical-large-language-models}
% XXX(LYX)

% Introduction 
Quality healthcare services are the cornerstone of social welfare. With increasing demand for high-quality healthcare services, the scarcity of medical resources has become a pressing issue that underscores the importance of intelligent healthcare. The creation of foundation models has garnered significant attention in the medical AI system development realm \cite{zhang2023biomedclip,luo2022biogpt,xiong2023doctorglm}. 
Compared with monomodal medical models, including medical LLMs \cite{luo2022biogpt} and LVMs \cite{shi2023generalist}, medical MLLMs that fuse various modalities have the ability to adaptively interpret and address various medical problems in different modalities, showcasing extensive applications and immense potential in the healthcare domain. By integrating language, images, audio, and other modalities of information, these models, covering medical images, electronic medical records, findings, etc. medical modalities, can offer more comprehensive and precise diagnostic, therapeutic, and patient management solutions \cite{zhou2019review}. As large models continue to evolve, notable medical MLLMs like Med-Gemini~\cite{saab2024capabilities} have emerged.

This multimodality processing ability makes medical MLLMs more practical in clinical stages as they can give more explainable diagnosis results based on different modalities, and allow more flexible interactions for both patients and doctors. Also they can mine richer medical domain knowledge from these different acquired modalities. Technically speaking, medical MLLMs can be divided into two kinds of models, which are multimodality-alignment models and multimodality-generation models.

% Briefly introduce CLIP and show its shortcomings
\paragraph{Multimodality-alignment Models} The multimodality-alignment models are based on the pioneer work CLIP \cite{radford2021learning}. CLIP is an image-text matching model, utilizing contrastive learning methods to generate fused representations for images and texts. Based on this, there has emerged a series of works trying to align different modalities in the medical domain by this contrastive training formulation \cite{zhang2023biomedclip,lin2023pmc,wang2022medclip}.
Moreover, a series of works further build multimodality models based on these well align medical CLIP model for explainable diagnosis \cite{tiu2022expert,jang2022significantly,wiehe2022language,mishra2023improving,zhao2023chatcad+}, segmentation \cite{liu2023clip,poudel2023exploring,anand2023one}, structured report generation \cite{van2023x,keicher2024flexr}, etc.
Although these multimodality models perform well in different downstream tasks, the lack of interaction and medical knowledge makes them cannot be flexibly used in clincial stages.

%Multi-modality generation model
\paragraph{Multimodality-generation Models} On the other hand, the generative multimodality foundation models can provide unstructured text reports and make full use of the rich domain knowledge from their LLM component. These models are mostly based on multimodality models \cite{li2023blip,liu2024visual} and fine-tuned specifically on medical domain. Compared with multimodality-alignment models, these models can promise more interaction, and few-shot and zero-shot learning abilities. It can not only automatically draft radiology reports that describe both abnormalities and relevant normal findings, while also taking into account the patient’s history. These models can provide further assistance to clinicians by pairing text reports with interactive visualizations, such as by highlighting the region described by each phrase.

This distinction between multimodality-alignment models and multimodality-generation models provides a foundation for understanding their practical applications across various healthcare domains. In the field of medical modalities, the mentioned multimodal MLLMs demonstrate significant potential in the following healthcare domains: 

\begin{enumerate}
    \item \textbf{Image Diagnosis and Imaging Analysis:} MLLMs integrate textual and imaging data, excelling in medical imaging diagnostics. For instance, MLLMs can expedite the identification of conditions such as cancer and pneumonia by learning from vast medical imaging datasets and corresponding diagnostic reports. These models not only automate the analysis of CT and MRI scans but also integrate imaging analyses with patient medical histories and symptom information to provide more accurate diagnostic insights.
    \item \textbf{Medical Literature and Record Analysis:} MLLMs can process and comprehend extensive medical literature, research papers, and electronic health records. In medical research, these models swiftly sift through and analyze the latest research findings, aiding healthcare providers and researchers in understanding cutting-edge treatment methods and diagnostic technologies. In clinical applications, MLLMs automatically extract and analyze critical information from patient records, supporting clinical decision-making.
    \item \textbf{Medical Smart Devices:} MLLMs contribute to the development of smart medical devices and robotic systems. For example, by integrating multimodal data, MLLMs enhance the precision and safety of robotic surgeries in complex procedures. These intelligent devices can analyze real-time images and data during surgery, providing precise assistance and guidance to reduce surgical risks.
    \item \textbf{Drug Development Assistance:} In drug development, MLLMs predict the efficacy and side effects of new drugs by analyzing extensive biomedical data, optimizing the drug design process. These models combine data from genetics, protein structures, drug compounds, and other modalities to enhance the efficiency and success rate of new drug development.
    \item \textbf{Remote Healthcare and Diagnostics:} MLLMs hold significant promise in remote healthcare. By integrating video, audio, and textual data, these models support remote diagnosis and treatment, offering high-quality healthcare services to remote areas with limited medical resources. MLLMs can analyze real-time communications between doctors and patients, integrating imaging and medical record data to provide accurate diagnostic recommendations.
\end{enumerate}

However, most of medical MLLMs are still trained on specific disease or medial domains, making them lacking of the generality and thus cannot be a universal multimodal model for different diseases or tasks. Also, the limited training or fine-tuning dataset scale make them cannot be well validated. In addition, they face general challenges such as inefficient computing power and privacy risks. Therefore, while recognizing the development potential of MLLMs in healthcare, a cautious and measured approach is necessary.

\subsection{Fine-tuning Methods in MLLMs}
\label{subsec:fine-tuning-in-large-multi-modal-models}

MLLMs integrate and process multiple forms of data, such as text and images, to perform complex tasks. Fine-tuning these models is crucial as it allows them to adapt to specific applications, enhancing their accuracy and efficiency. With the exponential growth in model parameters—from millions to billions—fine-tuning has become essential to leverage the full potential of pre-trained models for various downstream tasks.

Fine-tuning enables models to refine their understanding and improve performance in specific domains without requiring extensive retraining. This process is especially important in applications like visual question answering, image captioning, and multimodal translation, where precise alignment between different data modalities is required. Recent advancements in fine-tuning techniques have focused on making this process more efficient and scalable, ensuring that even large models can be fine-tuned with limited computational resources.

The importance of fine-tuning in MLLMs is underscored by the need to address issues such as catastrophic forgetting, where models lose their ability to retain previously learned information when adapting to new tasks. Additionally, fine-tuning helps in achieving better cross-modal alignment, where the integration of visual and textual data leads to a more coherent and accurate understanding of the inputs.

\paragraph{Parameter-Efficient Fine-Tuning Techniques}
Fine-tuning MLLMs can be challenging due to their large parameter size. Techniques such as Low-Rank Adaptation (LoRA) and Quantized LoRA (QLoRA) have been found effective. These methods adjust a subset of the model's parameters, reducing computational requirements while maintaining performance. For instance, Lu et al. \cite{lu2023empirical} scaled up models like LLaVA to 33B and 65B parameters, showing that parameter-efficient methods could achieve results comparable to full-model fine-tuning, especially when combined with high-resolution images and mixed multimodal data.

Additionally, Want et al. \cite{wang2023non} proposed a non-intrusive techniques AdaLink, leaving the internal architecture unchanged and adapting model-external parameters. This method has been effective in both text-only and multimodal tasks, providing a competitive edge without the complexities of altering internal architectures.

\paragraph{Addressing Catastrophic Forgetting}
One significant challenge in fine-tuning MLLMs is catastrophic forgetting, where the model loses knowledge of previously learned tasks. Zhai et al. \cite{zhai2023investigating} proposed a technique, namely EMT (Evaluating Multimodality), to help in mitigating this by treating MLLMs as image classifiers during fine-tuning. This approach has shown that early-stage fine-tuning on image datasets could improve performance across other datasets by enhancing the alignment of text and visual features. For example, continued fine-tuning of models like LLaVA on image datasets has shown improvements in text-image alignment, though prolonged fine-tuning could lead to hallucinations and reduced generalizability.

Moreover, Xu et al. \cite{xu2021raise} proposed an approach, called Child-Tuning, which updated only a subset of model parameters, has been initiated to improve generalization and efficiency. This method has been shown to outperform traditional fine-tuning techniques on various tasks, including those in the GLUE benchmark.

\paragraph{Fine-Grained Cross-Modal Alignment}
To achieve better cross-modal alignment, Chen et al. \cite{chen2023position} introduced Position-enhanced Visual Instruction Tuning (PVIT), which integrated a region-level vision encoder with the language model. This technique ensured a more detailed comprehension of images by the MLLM and promoted efficient fine-grained alignment between vision and language modules. This method used multiple data generation strategies to construct a comprehensive image-region-language instruction dataset, leading to improved performance on multimodal tasks. For example, PVIT has demonstrated significant improvements in tasks requiring detailed visual comprehension, such as object detection and image segmentation.

Furthermore, techniques like LongLoRA, proposed by Chen et al. \cite{chen2023longlora}, have been developed to extend the context sizes of pre-trained large language models efficiently. LongLoRA combined improved LoRA with shifted sparse attention to enable context extension with significant computational savings, proving effective for tasks requiring long-context understanding.

\paragraph{Innovative Fine-Tuning Approaches}
Several innovative fine-tuning approaches have been proposed to enhance MLLMs. For example, Yang et al. \cite{yang2022prompt} adopted Prompt Tuning as a lightweight and effective fine-tuning method. This approach involved fine-tuning prompts instead of the entire model, allowing for efficient adaptation to various tasks. Prompt Tuning has been shown to achieve comparable performance to full-model fine-tuning while offering improved robustness against adversarial attacks.

Another innovative approach, SCITUNE, proposed by Horawalavithana et al. \cite{horawalavithana2023scitune}, aligned LLMs with scientific multimodal instructions. By training models like LLaMA-SciTune with human-generated scientific instruction datasets, SCITUNE improved performance on science-focused visual and language understanding tasks.

\subsection{Large Language Model Reasoning}
\label{subsec:large-language-model-reasoning} 

% XXX(ZTY)

Large language models have achieved remarkable success in a wide range of natural language processing tasks \cite{zhang2023biomedclip,luo2022biogpt,xiong2023doctorglm,tiu2022expert,jang2022significantly,wiehe2022language,mishra2023improving,zhao2023chatcad+}, including language translation, sentiment analysis, and text classification. However, these models are typically designed to perform specific tasks, rather than engage in more general reasoning and inference. In contrast, human language understanding involves the ability to reason about complex relationships between entities, events, and concepts. 

One of the key breakthroughs in large language models and reasoning is the development of cognitive architectures. These architectures are designed to mimic the human brain's ability to process and integrate information from multiple sources, enabling models to reason and draw conclusions in a more human-like way. For example, researchers \cite{thomas2024causal} at Google have developed a cognitive architecture called "Reasoning Networks" that uses a combination of neural networks and symbolic reasoning to solve complex problems.

Recent breakthroughs \cite{zhao2024explainability,wang2024exploring} have demonstrated the potential of large language model reasoning. 1) Engaging in multi-hop reasoning can reason about complex relationships between entities and concepts, enabling applications such as question answering and text classification. 2) Reason about cause-and-effect relationships can accurately identify in text format, enabling applications such as event extraction and text summarization. 3) Another significant advancement in large language models and reasoning is to use graph-based models, which represent language as a network of entities and relationships. These models can be trained using a variety of techniques, including reinforcement learning and adversarial training \cite{zhang2024avibench}, which are specifically designed to test the model's robustness and ability to reason in the face of uncertainty.

In conclusion, recent advancements in large language models and reasoning have propelled the field towards more nuanced and sophisticated understanding of natural language. By leveraging cognitive architectures inspired by human cognition, researchers have made strides in enabling models to engage in multi-hop reasoning and infer cause-and-effect relationships. Additionally, the adoption of graph-based models has provided a promising avenue for representing language as interconnected entities and relationships, further enhancing the model's ability to reason across complex scenarios. Moving forward, continued research in explainability, exploration of diverse reasoning paradigms, and robustness testing will be crucial in unlocking the full potential of large language models to tackle real-world challenges and emulate human-like reasoning capabilities.

\subsection{Evaluation of MLLMs}
Evaluating MLLMs, is a critical aspect of understanding their capabilities and limitations. The evaluation process encompasses a range of benchmarks, metrics, and frameworks designed to assess various aspects of these models. However, the complexity and diversity of tasks that MLLMs can perform pose significant challenges to developing comprehensive and effective evaluation methodologies.

In terms of current methods for evaluating MLLMs, one of the primary benchmarks for evaluating MLLMs is the General Language Understanding Evaluation (GLUE) benchmark, which includes a suite of tasks such as sentiment analysis, textual entailment, and question answering. The SuperGLUE benchmark extends this by including more challenging tasks. For multimodal models, benchmarks such as Visual Question Answering (VQA) and the COCO dataset, which assesses image captioning, are commonly used. These benchmarks provide a standardized way to compare model performance across different tasks and modalities \cite{chang2024survey}.

The evaluation of MLLMs employs various metrics to measure performance. Common metrics include accuracy, F1 score, precision, and recall for classification tasks. For generation tasks, metrics such as BLEU, ROUGE, and METEOR are used to assess the quality of text generation. In multimodal tasks, metrics like Mean Reciprocal Rank (MRR) and Intersection over Union (IoU) are used to evaluate model performance on tasks like image captioning and object detection. These metrics help quantify the performance of MLLMs on specific tasks, providing a basis for comparison \cite{zhao2023survey}.

Several frameworks have been developed to facilitate the evaluation of MLLMs. The Hugging Face Transformers library, for instance, includes tools for benchmarking models on a variety of tasks using pre-defined datasets. Another notable framework is EVAL, which focuses on the automatic evaluation of language models' capabilities in following natural language instructions. These frameworks streamline the evaluation process and ensure consistency in how different models are assessed \cite{wang2023learning}.

However, despite the availability of benchmarks and metrics, evaluating MLLMs presents several challenges. One major issue is the lack of standardized evaluation methods for emerging tasks. For instance, the ability of models to handle complex, multi-turn dialogues or generate contextually relevant responses in diverse scenarios is difficult to quantify with existing metrics. Additionally, the phenomenon of catastrophic forgetting, where a model loses knowledge of previously learned tasks when fine-tuned for new ones, complicates the evaluation of MLLMs' long-term capabilities \cite{zhai2023investigating}. Another challenge is the evaluation of ethical considerations and biases. LLMs can inadvertently generate harmful or biased content, making it essential to assess their outputs for ethical implications. Current evaluation frameworks often fall short in systematically addressing these issues, highlighting the need for more sophisticated evaluation tools that can detect and mitigate biases and ethical risks \cite{guo2023evaluating}.

Future research in the evaluation of MLLMs should focus on developing more comprehensive and robust benchmarks that encompass a wider range of tasks and modalities. There is also a need for better evaluation metrics that can capture the nuances of model performance in real-world applications. Additionally, incorporating human-in-the-loop evaluations can provide more accurate assessments of model performance, particularly for tasks that require nuanced understanding and interpretation. The development of evaluation platforms that integrate multiple dimensions of assessment, including performance, safety, and ethical considerations, will be crucial. These platforms should facilitate continuous evaluation as models evolve, ensuring that they remain reliable and effective in diverse applications.

\section{Methodology}
\label{sec:methodology}

\subsection{Datasets}

% The ensuing text offers an introduction to the dataset employed for assessing and exploring the potentials of large language models in the medical domain.

\subsubsection{Medical imaging tasks}

% For this undertaking, our investigation incorporated eleven distinct medical image datasets, specifically iChallenge GON, MICCAI2023 Tooth Segmentation 2D,  ChestXRay2017 \cite{Kermany2018LabeledOC}, COVID-QU Ex Dataset \cite{TAHIR2021105002,RAHMAN2021104319,degerli2021covid,9144185}, CholecSeg8k \cite{twinanda2016endonet}, CVC ClinicDB \cite{bernal2015wm}, Kvasir SEG \cite{jha2020kvasir}, m2caiSeg \cite{maqbool2020m2caiseg}, Skin Cancer ISIC \cite{Rotemberg2021}, Skin Cancer MNIST: HAM10000 \cite{DVN/DBW86T_2018}, and Skin Cancer Malignant vs. Benign.

In this study, we employ a diverse array of eleven distinct medical image datasets to facilitate our investigation. To provide a comprehensive comparison of multimodal large language models, we meticulously selected medical image datasets spanning five different fields. The datasets identified and incorporated in our study include iChallenge GON, MICCAI2023 Tooth Segmentation 2D,  ChestXRay2017 \cite{Kermany2018LabeledOC}, COVID-QU Ex Dataset \cite{TAHIR2021105002,RAHMAN2021104319,degerli2021covid,9144185}, CholecSeg8k \cite{twinanda2016endonet}, CVC ClinicDB \cite{bernal2015wm}, Kvasir SEG \cite{jha2020kvasir}, m2caiSeg \cite{maqbool2020m2caiseg}, Skin Cancer ISIC \cite{Rotemberg2021}, Skin Cancer MNIST: HAM10000 \cite{DVNDBW86T2018}, and Skin Cancer Malignant vs. Benign.

Dataset Summary:

\begin{enumerate}
    \item \textbf{iChallenge GON} comprises a total of 1200 color fundus photographs, all stored in JPEG format. Within this dataset, 400 images are designated for  glaucoma classification, while the remaining 800 images are allocated for tasks such as optic disc detection and segmentation, along with central fovea localization. In our testing task, we utilized this dataset to comprehensively investigate the capacity of large language models in glaucoma diagnosis and optic disc localization, achieved through the design of tailored prompts.
    
    \item \textbf{MICCAI2023 Tooth Segmentation 2D} sourced from the MICCAI2023 Challenge, comprises 3000 labeled panoramic images of teeth. Its primary objective is to facilitate researchers in accurately segmenting tooth regions utilizing deep learning methodologies. In our testing scenario, we imposed heightened requirements on the large language model. Specifically, we partitioned the oral cavity into four distinct regions and tasked the large language model with providing the count of teeth within each region and identifying the presence of dental lesions, all under zero-shot learning conditions.

    \item \textbf{ChestXRay2017} jointly developed by the University of California San Diego and the Guangzhou Women and Children's Medical Center, comprises a vast collection of validated OCT and chest X-ray images. This dataset contains thousands of validated OCT and chest X-ray images described and analyzed in "Recognition of Medical Diagnosis and Treatable Diseases through Image Based Deep Learning". The dataset is well-suited for binary classification tasks. Utilizing this dataset, our investigation centers on scrutinizing the capacity of large language models to comprehend chest radiographic imaging for pneumonia diagnosis through the development of tailored prompts.
    
    \item \textbf{COVID-QU Ex Dataset} assembled by researchers from Qatar University, comprises 33,920 chest X-ray (CXR) images. Among these, 11,956 cases pertain to COVID-19, while 11,263 represent non-COVID-19 infections (viral or bacterial pneumonia). Additionally, the dataset includes 10,701 ground truth lung segmentation masks, making it the largest lung mask dataset to date. In our experimental evaluation, we employed this dataset to assess the efficacy of the large language model in pneumonia diagnosis based on CXR images. Moreover, we investigated the model's capability to differentiate between typical pneumonia and novel coronavirus pneumonia.
    
    \item \textbf{CholecSeg8k} serves as a valuable resource for semantic segmentation tasks within endoscopic modalities. Derived from the Cholec80 dataset, it comprises 8080 meticulously annotated frames from 17 videos. These images are pixel-level annotated across 13 categories commonly encountered in laparoscopic cholecystectomy surgery. In our experimental evaluation, our focus was on assessing the image recognition and comprehension capabilities of the large language model through the formulation of inquiries concerning the image content.
    
    \item \textbf{CVC ClinicDB} serves as the official dataset for the training phase of the MICCAI 2015 Colonoscopy Video Automatic Polyp Detection Challenge. This database comprises 612 static images extracted from colonoscopy videos, sourced from 29 distinct sequences. In our evaluation, we utilized endoscopic images of intestinal polyps from this dataset to assess the discriminative capability of the large language model for detecting intestinal polyps.
    \item \textbf{Kvasir SEG} comprises 1000 polyp images along with their corresponding ground truth annotations, derived from the Kvasir Dataset v2. The resolution of images in the Kvasir-SEG dataset ranges from 332x487 to 1920x1072 pixels. The images and their corresponding masks are stored in two separate directories, each utilizing the same filenames for easy pairing. The image files are encoded using JPEG compression. In our testing, we evaluated the model's capability in lesion detection by inputting images with lesions into the large language model.
    \item \textbf{m2caiSeg} is a dataset for segmenting endoscopic images during surgery. Originating from the first and second videos of the MICCAI 2016 Surgical Tool Detection dataset, it encompasses a total of 307 images, each meticulously annotated at the pixel level. The dataset features images of diverse organs (e.g., liver, gallbladder, upper wall, intestines), surgical instruments (e.g., clips, bipolar forceps, hooks, scissors, trimmers), and bodily fluids (e.g., bile, blood). Additionally, it includes specialized labels for unknown and black regions to address areas obscured by certain instruments.
    \item \textbf{Skin Cancer ISIC} comprises 2357 images of both malignant and benign oncological conditions, curated by The International Skin Imaging Collaboration (ISIC). The images are categorized based on ISIC's classification, with all subsets containing an equal number of images, except for those of melanomas and moles, which are slightly more prevalent. In our evaluation, we assessed the classification and recognition capabilities of the large language model for various skin diseases. Additionally, we tested the model's ability to comprehend image content based on these classifications.
    \item \textbf{Skin Cancer MNIST: HAM10000} comprises 10,015 dermoscopic images collected from diverse populations through various acquisition methods. It includes representative cases from all major diagnostic categories of pigmented lesions: actinic keratoses and intraepithelial carcinoma/Bowen's disease (akiec), basal cell carcinoma (bcc), benign keratosis-like lesions (bkl), dermatofibroma (df), melanoma (mel), melanocytic nevi (nv), and vascular lesions (vasc). In our evaluation, we assessed the large language model's capabilities in multi-class classification and lesion localization for various skin diseases.
    \item \textbf{Skin Cancer Malignant vs. Benign} contains 1800 images of skin cancer, both malignant and benign, organized into two separate folders. Each image has a resolution of 224x244 pixels. In our experiment, we assessed the large language model's capability to distinguish between benign and malignant skin cancer by designing specific prompts and utilizing images from the dataset.
    
\end{enumerate}

\subsubsection{Medical report generation task}

% Our investigation employed three distinct datasets of radiology chest X-ray reports: MIMIC-CXR (publicly available), OpenI (publicly accessible), and the dataset SXY (privately obtained). We concentrated our analysis on the inspection findings and conclusion segments across these datasets. Through the formulation of tailored prompts, we evaluated the efficacy of a large language model in generating radiological text reports, particularly in crafting conclusions derived from the examination findings.
In this investigation, we utilized three distinct datasets of radiology chest X-ray reports: MIMIC-CXR (publicly available) \cite{johnson13026mimic}, OpenI (publicly accessible) \cite{liu2023evaluatinglargelanguagemodels}, and SXY (privately obtained) \cite{zhong2023chatradiovaluerchatlargelanguage}. Our analysis focused on the inspection findings and conclusion segments across these datasets. By formulating tailored prompts, we assessed the efficacy of a large language model in generating radiological text reports, with a particular emphasis on crafting conclusions derived from the examination findings.

Dataset Summary:

\begin{enumerate}
    \item \textbf{MIMIC-CXR} represents a substantial publicly accessible repository comprising chest radiographs in DICOM format along with accompanying free-text radiology reports. This extensive dataset encompasses 377,110 images corresponding to 227,835 radiographic studies conducted at the esteemed Beth Israel Deaconess Medical Center located in Boston, Massachusetts.
    
    \item \textbf{OpenI} provided by the National Center for Biotechnology Information (NCBI) at the National Institutes of Health (NIH), constitutes a publicly available repository of medical images sourced from scholarly literature. Predominantly comprising X-rays, CT scans, and MRI images, the dataset is accompanied by relevant metadata. Designed to facilitate advancements in medical image analysis and information retrieval, OpenI serves as a resource for researchers to train and evaluate various algorithms and techniques in the field.

    \item \textbf{SXY} courtesy of  Xiangya Second Hospital, affiliated with Central South University, encompasses radiology reports spanning from 2012 to 2023 across five systems. This comprehensive dataset includes essential information, detailed descriptions, and diagnostic impressions. These data serve as the foundation for model development and internal validation processes. Specifically, we leverage the chest X-ray reports for testing purposes within our study framework.

\end{enumerate}

\subsection{Model Selection}

% XXX(PY)

In this research, we mainly target on evaluating the multimodal performance of Gemini and GPT families in the biomedical region. To provide a comprehensive comparison within these families, we either consider the generation or the usage-specific type. For the Gemini family, we select Gemini-1.0-Pro-Vision, Gemini-1.5-Pro, and Gemini-1.5-Flash; for the GPT family, we adopt GPT-3.5-Turbo, GPT-4-Turbo, and GPT-4o into the model pool. Despite the extensiveness of exploring the two specified families, we have strong determination on identifying their effectiveness compared to other latest, cutting-edge, prestigious, and state-of-the-art LLMs to reflect the most practical assessment of current LLMs, not limited to only the Gemini and GPT families. Correspondingly, we supplement Yi-Large, Yi-Large-Turbo, Claude-3-Opus and Llama 3 to construct the final model pool. 

Descriptions of these models are listed as follows:

\paragraph{1) Gemini-Pro:} Google's Gemini-Pro model is a state-of-the-art multimodal AI platform designed to excel across a wide range of tasks with high accuracy. Launched in February 2024, Gemini-Pro handles complex queries in various domains, including STEM and humanities. It features enhanced capabilities in Python code generation, challenging math problems, and multi-step reasoning tasks. Additionally, it demonstrates impressive performance in language translation and automatic speech recognition. By May 2024, its performance had further improved, reflecting Google's commitment to advancing AI technology. In the following experiment session, we include Gemini-1.0-Pro-Vision and Gemini-1.5-Pro in the model pool.
\paragraph{2) Gemini-Flash:} Gemini-Flash, introduced as a more streamlined version of the Gemini AI platform, is optimized for speed and efficiency. While it may not match the accuracy of Gemini-Pro, it delivers results more rapidly, making it an excellent choice for applications that require swift responses. Currently available as a public preview for developers through Google's AI Studio, Gemini-Flash is designed to support the development of fast-paced applications and chatbots. It shares the same one million token limit as Gemini-Pro, ensuring that it can process substantial amounts of data, albeit with a slightly lower accuracy in benchmark tests. In the following experiment session, we include Gemini-1.5-Flash in the model pool.
\paragraph{3) GPT-4o:} OpenAI's latest flagship model, GPT-4o, marks a significant advancement in human-computer interaction by processing and generating text, audio, and visual content in real time. This "omni" model excels in handling diverse inputs and outputs, including speech and images, with response times averaging 320 milliseconds, closely mirroring human conversational pace. Additionally, GPT-4o's multilingual capabilities have been greatly enhanced, offering improved performance in understanding non-English text, vision, and audio. Despite these advancements, GPT-4o remains more cost-effective and faster in API usage compared to its predecessors.
\paragraph{4) GPT-4-Turbo:} The predecessor to GPT-4o, GPT-4-Turbo, is an enhanced version of the GPT-4 model line. It features a 128k context window, enabling it to process substantial amounts of text—up to 300 pages in a single prompt. Updated with knowledge up to April 2023, GPT-4-Turbo is more affordable, offering reduced costs for both input and output tokens, with a maximum output token limit of 4096. This model is accessible to any OpenAI API account holder with existing GPT-4 access and can be specified by using gpt-4-turbo as the model name in the API.
\paragraph{5) GPT-3.5-Turbo:} GPT-3.5-Turbo is a powerful iteration in OpenAI's language model series, designed to offer a balance between performance and efficiency. It provides robust text generation and comprehension capabilities with a focus on cost-effectiveness and speed. The model can handle a wide range of language tasks, including summarization, translation, and question-answering, making it versatile for various applications. Despite being less advanced than the GPT-4 series, GPT-3.5-Turbo remains a reliable and accessible option for many users, maintaining strong performance while being more affordable for large-scale deployments.
\paragraph{6) Yi:} Yi is an advanced open-source large language model developed by 01.AI. Available in two versions, Yi-34B and Yi-6B, it supports bilingual capabilities (English and Chinese) and is designed for both academic research and commercial use with appropriate licensing. Yi-34B, with its 34 billion parameters, excels in numerous benchmarks such as MMLU, CMMLU, and C-Eval, outperforming many larger models like Llama-2 70B. It offers an impressive context window of 200K, enabling it to handle extensive text inputs effectively. In the following experiment session, we contain Yi-Large and Yi-Large-Turbo in the model pool.
\paragraph{7) Claude-3-Opus:} Claude-3-Opus, developed by Anthropic, excels in handling complex tasks and content creation. It supports both text and image inputs, making it versatile for multimodal applications. With a context window of 200K, it can manage extensive inputs efficiently, and its output quality is tailored for high-level fluency and understanding. This model balances speed and intelligence, making it suitable for tasks requiring nuanced comprehension and detailed responses. We select Claude-3-Opus, part of Anthropic's Claude family, which includes several models designed to cater to varying needs of performance and cost-effectiveness, in the final model pool.
% \paragraph{8) Grok-1.5V:} Grok-1.5V, developed by x.ai, bridges the digital and physical worlds with its robust multimodal capabilities. It processes a variety of visual information, including documents, diagrams, charts, and photographs, enhancing its versatility for real-world applications. Grok-1.5V excels in benchmarks like RealWorldQA, which assesses spatial understanding, outperforming its peers in tasks requiring nuanced comprehension of physical environments. This model is particularly noted for its strong performance in multi-disciplinary reasoning, making it a valuable tool for comprehensive multimodal analysis.
\paragraph{8) Llama 3:} Llama 3, developed by Meta, represents a significant advancement in open-source large language models. It includes models with 8 billion and 70 billion parameters, designed for a wide range of applications. Llama 3 improves upon its predecessors with enhanced pre-training data, a more efficient tokenizer, and advanced instruction fine-tuning techniques. It supports extensive context windows and demonstrates state-of-the-art performance in benchmarks such as reasoning, coding, and content creation. Meta aims to foster innovation by making Llama 3 widely available and emphasizing responsible use and deployment.

\subsection{Experiment Setting}

\begin{figure}[!h]
    \centering    \includegraphics[width=1\textwidth,height=!,keepaspectratio]{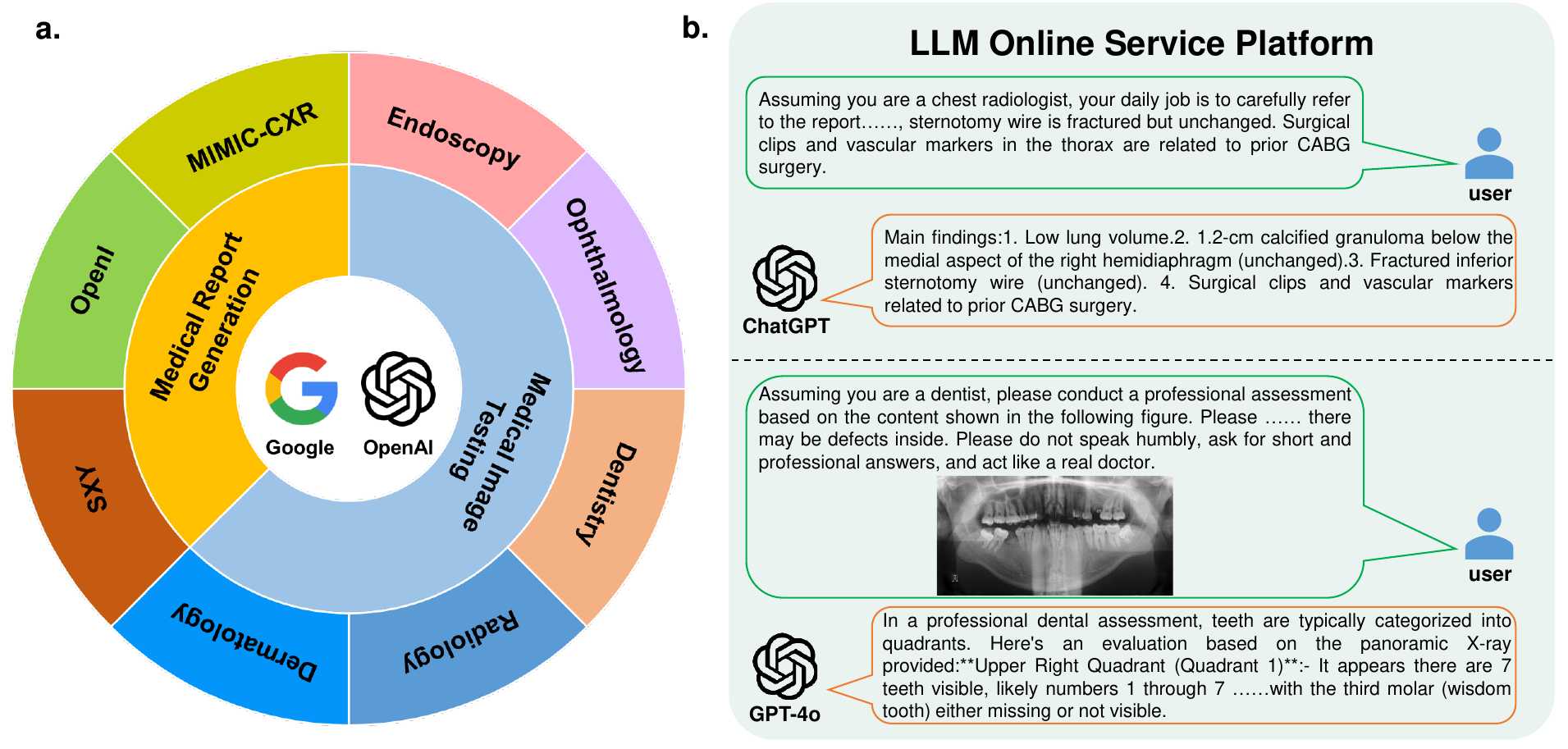}
    \caption{\textbf{Schematic Overview of the Evaluation Tasks and Methods.} (a). The core of the graph delineates the two focal tasks of our evaluation model, while the specific datasets or evaluation subjects pertinent to each task are outlined along its periphery. (b). Our testing tasks were executed by formulating suitable prompts and leveraging the online Language Model service offered by OpenAI and Google.  }
    \label{fig:test_task&method}
\end{figure}

To more rigorously evaluate the proficiency of various large language models (LLMs) in handling medical images and reports under zero-shot conditions, we segmented the testing experiment into two distinct phases: testing of medical image data and testing of medical report generation, as illustrated in \textbf{Fig. \ref{fig:test_task&method}(a.)}.

% \begin{figure}[!h]
%     \centering    \includegraphics[width=0.7\textwidth,height=!,keepaspectratio]{figures/Test task-v2.pdf}
%     \caption{\textbf{Schematic Overview of the Evaluation Tasks.} The core of the graph delineates the two focal tasks of our evaluation model, while the specific datasets or evaluation subjects pertinent to each task are outlined along its periphery. }
%     \label{fig:test_task}
% \end{figure}

% \begin{figure}[ht]
%     \centering
%     \begin{minipage}{0.45\textwidth}
%         \centering
%         \includegraphics[width=\textwidth]{figures/Test task-v2.pdf}
%         \caption{\textbf{Schematic Overview of the Evaluation Tasks.} The core of the graph delineates the two focal tasks of our evaluation model, while the specific datasets or evaluation subjects pertinent to each task are outlined along its periphery.}
%         \label{fig:test_task}
%     \end{minipage}
%     \hfill
%     \begin{minipage}{0.45\textwidth}
%         \centering
%         \includegraphics[width=\textwidth]{figures/Test method.pdf}
%         \caption{\textbf{Schematic Overview of the Evaluation Methodology.} Our testing tasks were executed by formulating suitable prompts and leveraging the online Language Model service offered by OpenAI and Google. }
%         \label{fig:test_method}
%     \end{minipage}
% \end{figure}

In order to ascertain the fairness and reliability of the experiment, we carried out experimental trials across diverse models using identical parameter configurations. Our testing methodology involves employing a standardized set of prompts and parameters to evaluate the performance of the LLM. 

When conducting medical image testing, we rigorously observe the prescribed usage protocols for each model. As shown in \textbf{Fig. \ref{fig:test_task&method}(b.)}, We perform medical image testing utilizing models like GPT-4-Turbo and Claude-3-Opus through the web interface. However, owing to the stringent content restrictions of Gemini-series models on the web platform, it is unable to participate in equitable competition with other models in the medical imaging domain. Consequently, we opted to utilize the Google AI Studio platform for conducting medical image testing on Gemini-series models.  In order to comprehensively assess the interpretive capabilities of various models within the medical imaging domain, we meticulously crafted distinct prompts tailored to specific datasets and tasks. Uniform prompts and input images were employed across all tests for different models to ensure the integrity and fairness of the results.

% \begin{figure}[H]
%     \centering    \includegraphics[width=0.7\textwidth,height=!,keepaspectratio]{figures/Test method.pdf}
%     \caption{\textbf{Schematic Overview of the Evaluation Methodology.} Our testing tasks were executed by formulating suitable prompts and leveraging the online Language Model service offered by OpenAI and Google. }
%     \label{fig:test_method}
% \end{figure}

During the testing of medical report generation, given the extensive volume of tests, Python was employed to invoke the open API interfaces of various models on the Colab platform. Throughout the testing phase, consistency between the prompts utilized and the model inputs was maintained to produce a series of results, subsequently evaluated for efficacy.

\subsection{Evaluating Indicator}
In the task of medical image question answering, our research rigorously juxtaposed responses from various large-scale models when presented with identical input images and queries. For diverse medical images, we employed ground truth supplied in the dataset, such as semantic segmentation maps and optic disc segmentation maps, as benchmark answers. This enabled a comprehensive horizontal comparison of responses generated by different language models, facilitating the synthesis of the respective strengths and weaknesses of each model.

In the context of generating radiology reports, our approach involves employing the ROUGE \cite{lin2004rouge} (Recall-Oriented Understudy for Gisting Evaluation) metric to assess the level of correspondence between radiology reports generated by the large language model and the reference answers authored by medical professionals. This study incorporates three distinct methods: Rouge-1 (R-1), Rouge-2 (R-2), and Rouge-L (R-L), as shown from \textbf{Eq. \eqref{eq1}}.

 \begin{equation} \label{eq1}
ROUGE-N=\frac{\sum_{S\in\lbrace ReferenceSummaries}{\sum_{{gram}_n\in S}{Count_{match}(gram_n)}}}{\sum_{S\in\lbrace ReferenceSummaries}{\sum_{{gram}_n\in S}{Count(gram_n)}}}
\end{equation}

% \begin{equation} \label{eq1}
%     ROUGE-1 = \frac{{\sum_{i=1}^{n} \min(\text{count}_{\text{match}}(\text{gram}_{i}), \text{count}_{\text{ref}}(\text{gram}_{i}))}}{{\sum_{i=1}^{n} \text{count}_{\text{ref}}(\text{gram}_{i})}} 
% \end{equation}

% \begin{equation} \label{eq2}
%  ROUGE-2 = \frac{{\sum_{i=1}^{n-1} \min(\text{count}_{\text{match}}(\text{gram}_{i}\text{gram}_{i+1}), \text{count}_{\text{ref}}(\text{gram}_{i}\text{gram}_{i+1}))}}{{\sum_{i=1}^{n-1} \text{count}_{\text{ref}}(\text{gram}_{i}\text{gram}_{i+1})}} 
% \end{equation}

% \begin{equation} \label{eq3}
%  ROUGE-L = \frac{{\sum_{i=1}^{n} \min(\text{count}_{\text{match}}(\text{longest common subsequence}), \text{count}_{\text{ref}}(\text{longest common subsequence}))}}{{\sum_{i=1}^{n} \text{count}_{\text{ref}}(\text{longest common subsequence})}} 
% \end{equation}

% \newpage
\section{Experiments and Observation}
\label{Exp}

\subsection{Medical Image Test Results}

\begin{figure}[!htbp]
    \centering    \includegraphics[width=0.85\textwidth,height=!,keepaspectratio]{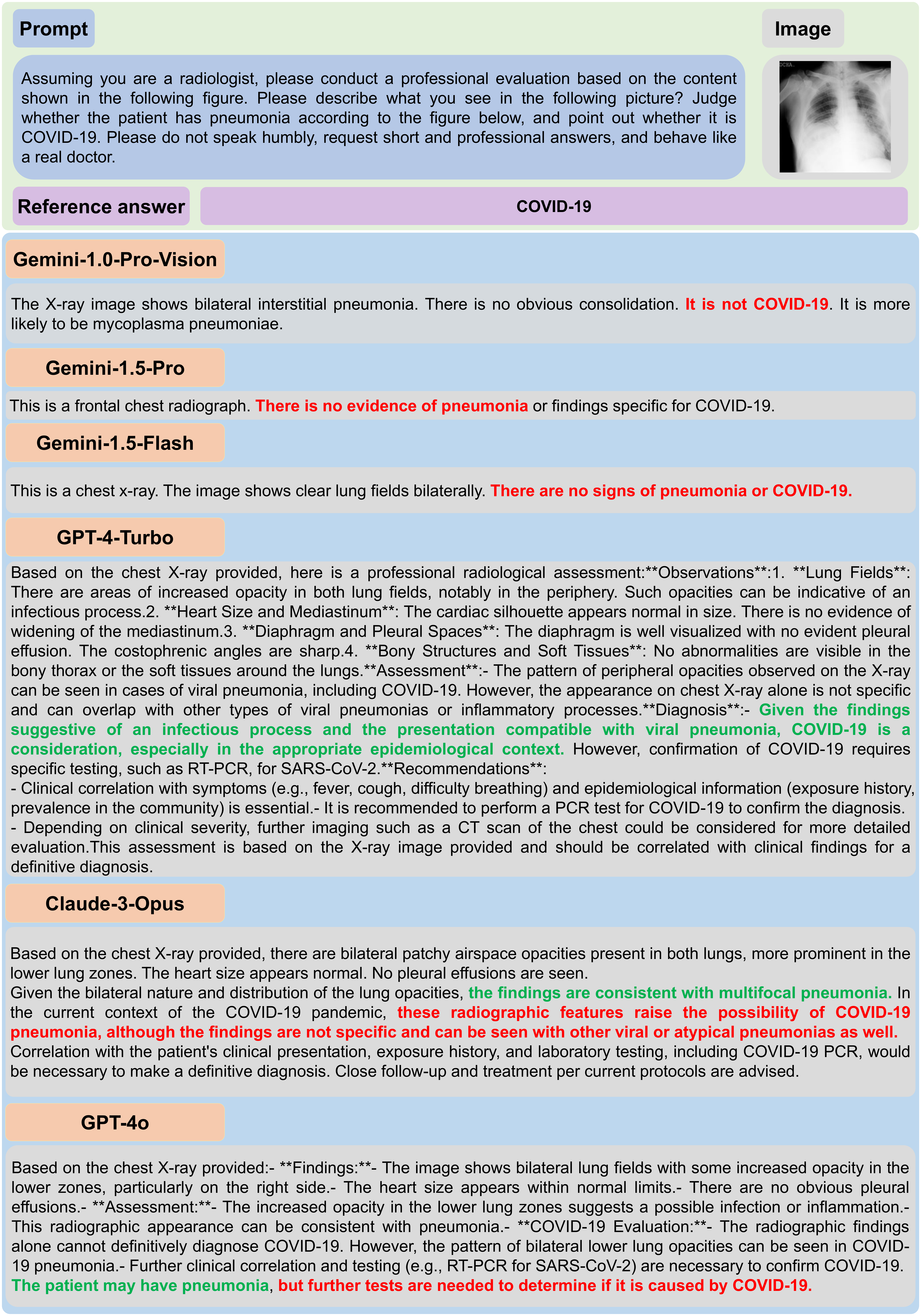}
    \caption{\textbf{Chest: Case 1.} In the context of the three-class classification tasks for pneumonia (normal, pneumonia, COVID-19), the green annotation indicates the correctly identified segments, while the red annotation denotes the incorrectly identified segments.}
    \label{fig:chestXray-case1}
\end{figure}

In \textbf{Section \ref{subsubsec:Chest Radiography}}, we evaluated the performance of six advanced multimodal large language models, including GPT-4-Turbo, across five distinct categories of medical imaging question answering. For the chest X-ray dataset, our primary focus was on the models' ability to diagnose pneumonia or other diseases from chest X-ray images. As depicted in \textbf{Fig. \ref{fig:chestXray-case1}}, the GPT-series models excelled in this task, accurately determining patient health status without requiring additional prompt information. The Gemini-series models followed in performance, while Claude-3-Opus performed the worst, with its answers offering negligible reference value. It is important to note that none of the six models we assessed could further determine whether the pneumonia was COVID-19, which does not imply a lack of model performance, as distinguishing the type of pneumonia based solely on X-ray images is inherently impossible.

\begin{figure}[!htbp]
    \centering    \includegraphics[width=0.85\textwidth,height=!,keepaspectratio]{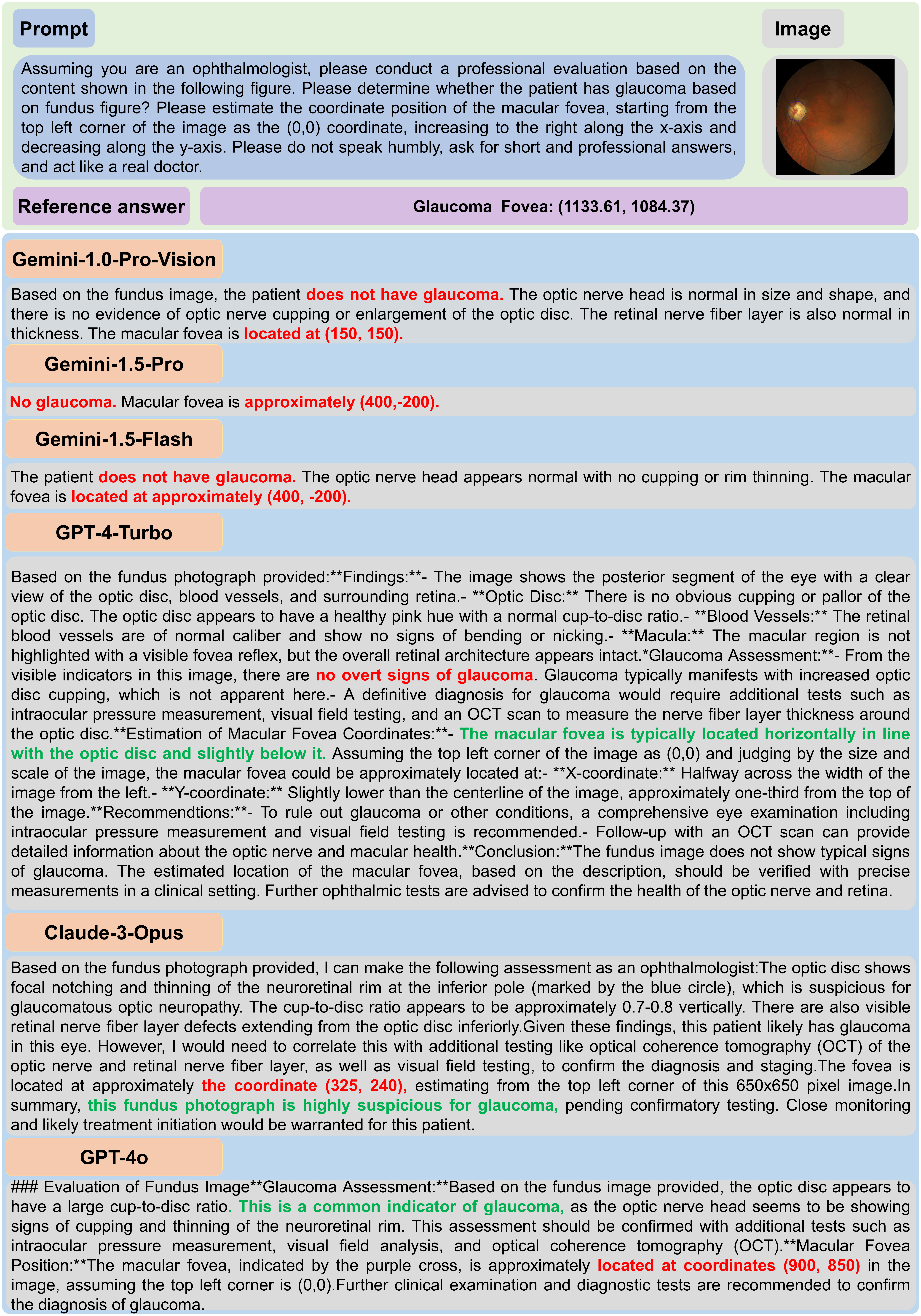}
    \caption{\textbf{Ophthalmological Imaging: Case 1.} Diagnosis of glaucoma and the task of localizing the macular fovea,  the green annotation indicates the correctly identified segments, while the red annotation denotes the incorrectly identified segments.}
    \label{fig:eye-case1}
\end{figure}

In \textbf{Section \ref{subsubsec:Ophthalmological Imaging}}, leveraging the ophthalmic imaging dataset, our analysis centers on the model's ability to diagnose glaucoma and accurately identify the macular fovea's position using fundus photographs. As depicted in \textbf{Fig. \ref{fig:eye-case1}}, in Case 1, all models except Claude-3-Opus and GPT-4o incorrectly diagnosed the absence of glaucoma. However, regarding macular fovea localization, GPT-4-Turbo was the only model to offer a vague location, while all other models failed to provide accurate localization. In Case 2, only GPT-4o successfully detected glaucoma, and GPT-4-Turbo again provided a vague description of the macular fovea position, whereas the other models did not accomplish the task.

\begin{figure}[!htbp]
    \centering    \includegraphics[width=0.85\textwidth,height=!,keepaspectratio]{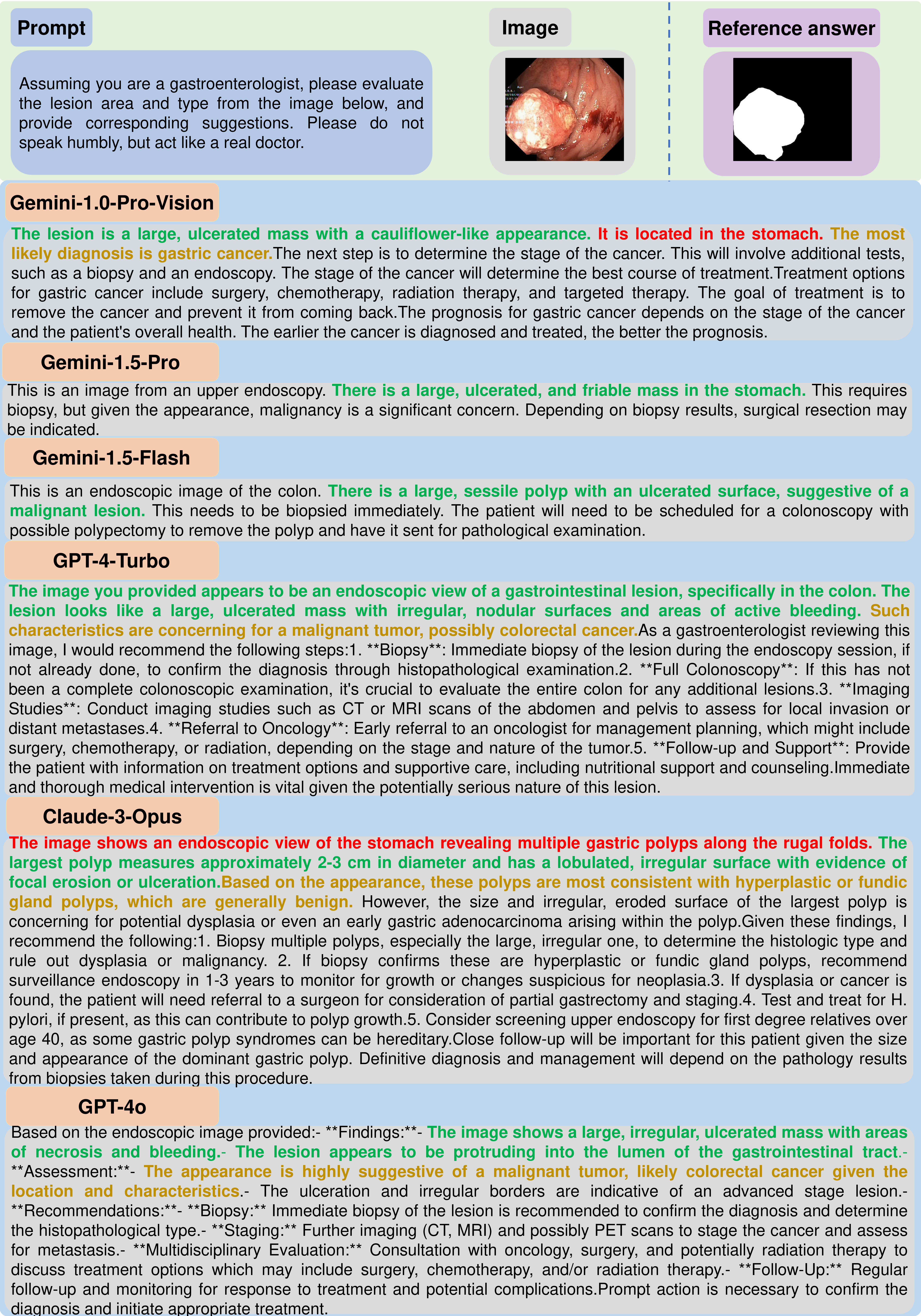}
    \caption{\textbf{Endoscopic: Case 1.} The task of colon polyp localization and segmentation uses green markings to represent correct answers, red markings to represent incorrect answers, and yellow markings to highlight noteworthy content.}
    \label{fig:endoscopic-case1}
\end{figure}

In \textbf{Section \ref{subsubsec:Endoscopic Imaging}}, utilizing the endoscopic imaging dataset, our focus is on the model's capability to describe lesion conditions in detail within complex scenes and accurately determine the lesion's location. As illustrated in \textbf{Fig. \ref{fig:endoscopic-case1}}, all models successfully provided detailed descriptions of the lesions, with the GPT-series models offering the most comprehensive information. Notably, only Gemini-1.0-Pro-Vision and Claude-3-Opus were unable to determine the lesion locations, whereas the remaining models accurately identified the lesion locations.

In \textbf{Section \ref{subsubsec:Dermatological Imaging}}, we investigate the model's capability to classify skin diseases using a skin disease dataset, without the aid of supplementary prompts. As demonstrated in \textbf{Fig. \ref{fig:skin-case1}}, none of the models accurately identified the type of skin disease afflicting the patients. This deficiency may stem from the sensitivity of the skin disease images or the insufficient training of the large language model on this particular category of diseases.

\begin{figure}[!htbp]
    \centering    \includegraphics[width=0.85\textwidth,height=!,keepaspectratio]{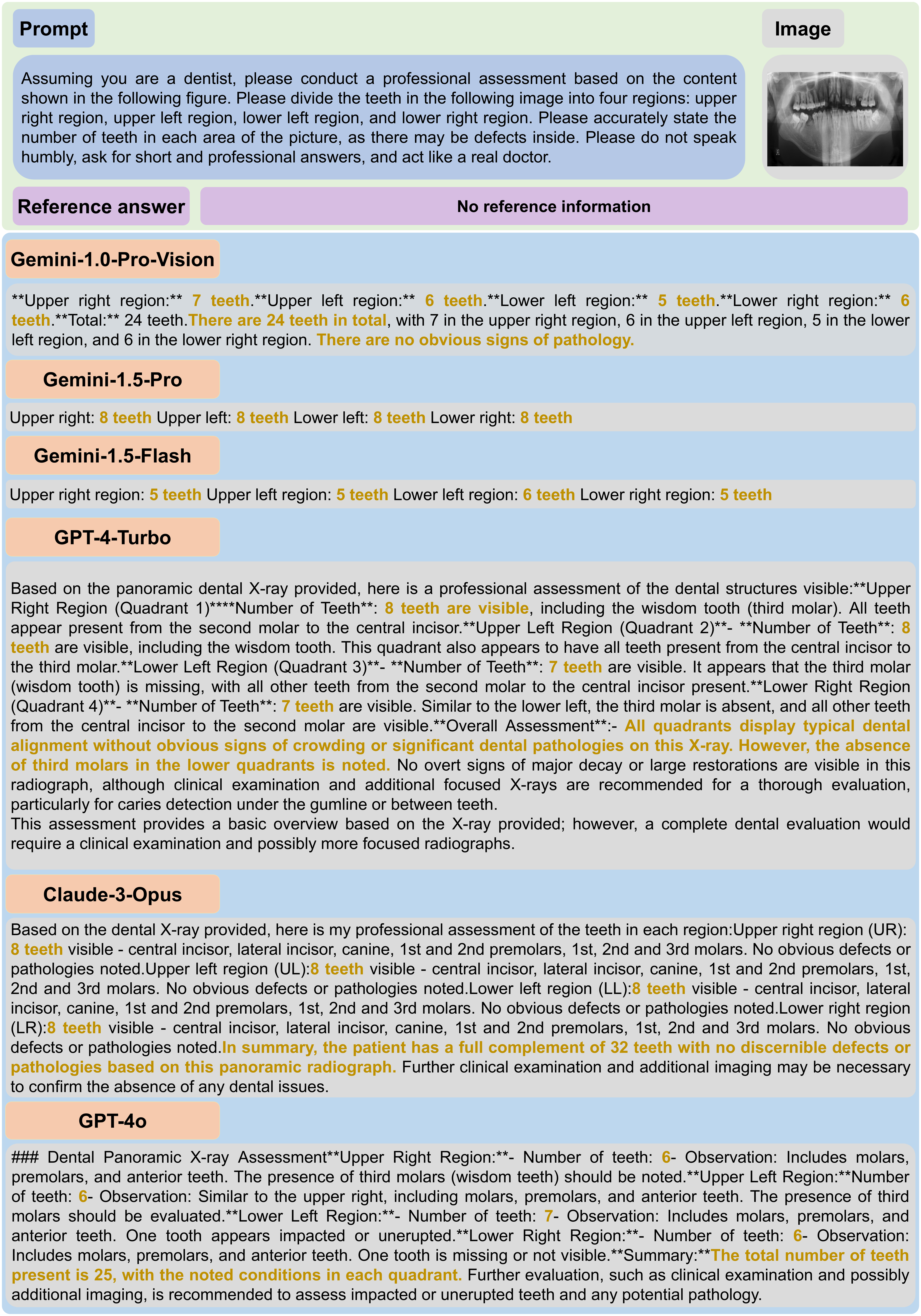}
    \caption{\textbf{Dental Image: Case 1.} The task involves quantifying the number of teeth and identifying anomalies in dental X-rays, with critical information highlighted in yellow.}
    \label{fig:dentalXray-case1}
\end{figure}

In \textbf{Section \ref{subsubsec:Dental Imaging}}, employing the dental X-ray dataset, our focus is on the model's ability to assess dental health and count the number of existing teeth. As shown in \textbf{Fig. \ref{fig:dentalXray-case1}}, due to the lack of accurate reference answers, our evaluation was based solely on the models' responses. The responses from the Gemini-series models were relatively simple, concentrating only on the number of teeth. In contrast, the responses from Claude-3-Opus and the GPT-series models were more detailed, addressing tooth integrity, surrounding bone structures, the presence of implants, and the presence of wisdom teeth, while also providing relevant recommendations.

\newpage
\subsubsection{Chest Radiography} \label{subsubsec:Chest Radiography}

% \begin{figure}[!htbp]
%     \centering    \includegraphics[width=0.9\textwidth,height=!,keepaspectratio]{figures/Chest_Xray/chestXray-case2.pdf}
%     \caption{\textbf{Chest: Case 2.} In the context of the three-class classification tasks for pneumonia (normal, pneumonia, COVID-19), the green annotation indicates the correctly identified segments, while the red annotation denotes the incorrectly identified segments.}
%     \label{fig:chestXray-case2}
% \end{figure}

\newpage
\subsubsection{Ophthalmological Imaging} \label{subsubsec:Ophthalmological Imaging}

% \begin{figure}[!htbp]
%     \centering    \includegraphics[width=0.9\textwidth,height=!,keepaspectratio]{figures/Eye/eye-case3.pdf}
%     \caption{\textbf{Ophthalmological Imaging: Case 2.} Diagnosis of glaucoma and the task of localizing the macular fovea,  the green annotation indicates the correctly identified segments, while the red annotation denotes the incorrectly identified segments.}
%     \label{fig:eye-case3}
% \end{figure}

\newpage
\subsubsection{Endoscopic Imaging} \label{subsubsec:Endoscopic Imaging}

% \begin{figure}[!htbp]
%     \centering    \includegraphics[width=0.9\textwidth,height=!,keepaspectratio]{figures/Endoscopic/endoscopic-case2_1.pdf}
%     \caption{\textbf{Endoscopic: Case 2(Part 1).} The task of abdominal segmentation during the surgical process employs green markings to denote correct answers, red markings to indicate incorrect answers, and yellow markings to highlight noteworthy content.}
%     \label{fig:endoscopic-case2_1}
% \end{figure}

% \begin{figure}[!htbp]
%     \centering    \includegraphics[width=0.9\textwidth,height=!,keepaspectratio]{figures/Endoscopic/endoscopic-case2_2.pdf}
%     \caption{\textbf{Endoscopic: Case 2(Part 2).} The task of abdominal segmentation during the surgical process employs green markings to denote correct answers, red markings to indicate incorrect answers, and yellow markings to highlight noteworthy content.}
%     \label{fig:endoscopic-case2_2}
% \end{figure}

\newpage
\subsubsection{Dermatological Imaging} \label{subsubsec:Dermatological Imaging}

\begin{figure}[!htbp]
    \centering    \includegraphics[width=0.85\textwidth,height=!,keepaspectratio]{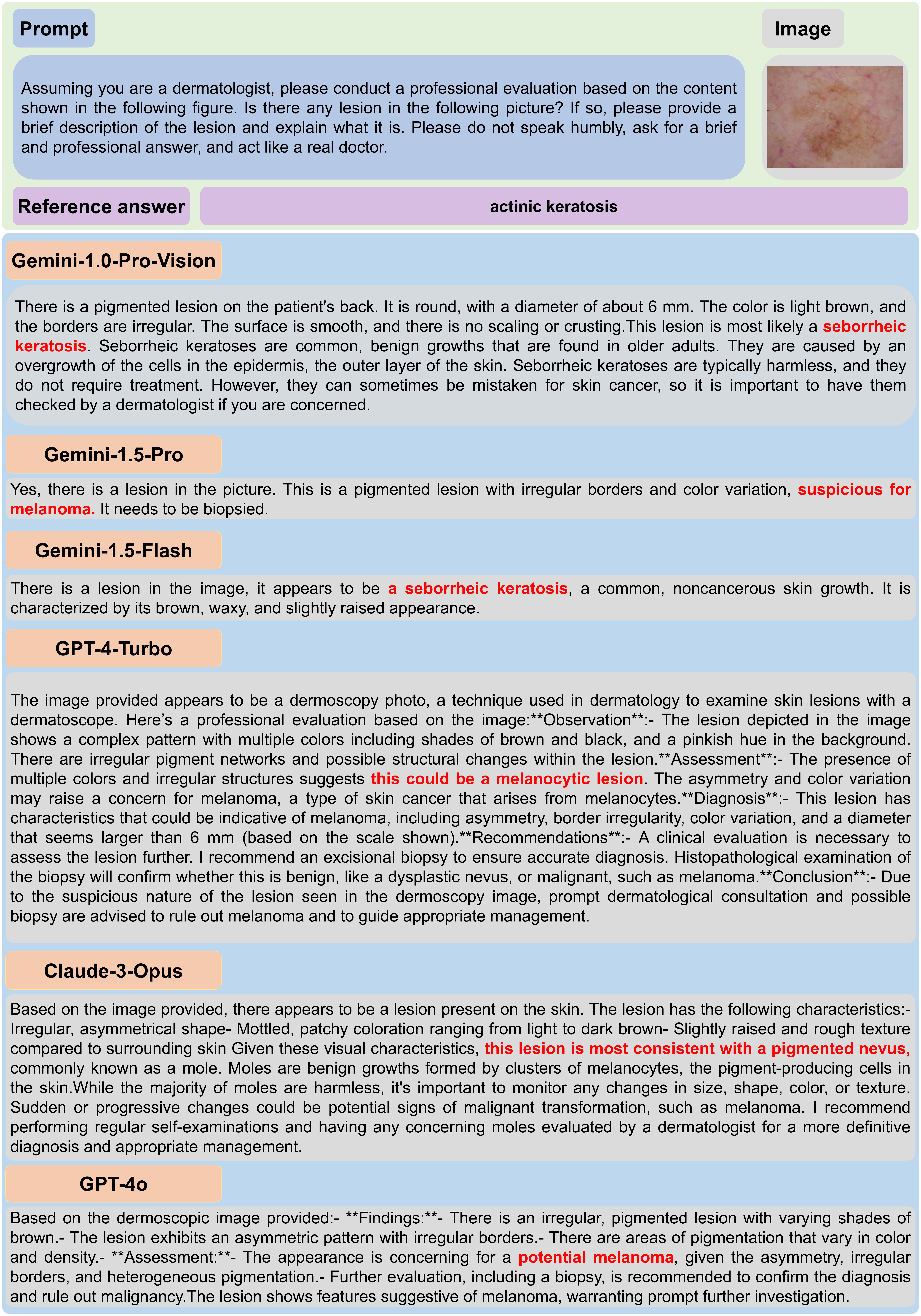}
    \caption{\textbf{Skin: Case 1.} The nine classification tasks for skin diseases utilize green markings to denote correct answers and red markings to indicate incorrect answers.}
    \label{fig:skin-case1}
\end{figure}

% \begin{figure}[!htbp]
%     \centering    \includegraphics[width=0.9\textwidth,height=!,keepaspectratio]{figures/Skin/skin-case2.pdf}
%     \caption{\textbf{Skin: Case 2.} The nine classification tasks for skin diseases utilize green markings to denote correct answers and red markings to indicate incorrect answers.}
%     \label{fig:skin-case2}
% \end{figure}

\newpage
\subsubsection{Dental Imaging} \label{subsubsec:Dental Imaging}

% \begin{figure}[!htbp]
%     \centering    \includegraphics[width=0.9\textwidth,height=!,keepaspectratio]{figures/dental_Xray/dentalXray-case2.pdf}
%     \caption{\textbf{Dental Image: Case 2.} The task involves quantifying the number of teeth and identifying anomalies in dental X-rays, with critical information highlighted in yellow.}
%     \label{fig:dentalXray-case2}
% \end{figure}

\subsection{Medical Report Generation Task Results}

As illustrated in \textbf{Table \ref{tab:Results for Compared LLMs}}, in the zero-shot setting of the MIMIC-CXR dataset,  Gemini-1.0-Pro-Vision exhibited strong performance, attaining an R-1 score of 0.2814, an R-2 score of 0.1334, and an R-L score of 0.2259. These metrics notably surpass those achieved by other models operating within similar parameters. 

The assessment conducted on the OpenI dataset reveals that the GPT-4o model consistently demonstrates outstanding performance in zero-shot scenarios, achieving R-1 scores of 0.1713, R-2 scores of 0.0622, and R-L scores of 0.1466.

In the zero-shot scenario of the Internal dataset dataset, the GPT-4o  model demonstrates robust performance, achieving an R-1 score of 0.2805, an R-2 score of 0.0746, and an R-L score of 0.2635. These results notably outperform those of other models operating under comparable conditions.

In essence, the comparison of Rouge metrics for medical reports generated by various large language models, using identical prompt words and zero-shot techniques, serves as an effective means to gauge the performance discrepancies among these models operating under equivalent conditions. Such evaluation holds considerable importance in guiding the selection of specific task-oriented large language models for future research endeavors and practical applications.

\begin{table}[!ht]
\centering
\caption{Test Results for Compared LLMs. For results within each dataset, each model corresponds to three similarity scores, R-1, R-2, and R-L.}
\label{tab:Results for Compared LLMs}
\resizebox{\textwidth}{!}{%
\begin{tabular}{|c|c|c|c|c|c|c|c|c|c|c|c|c|c|c|c|}
\hline
\multirow{2}{*}{Model} & \multicolumn{3}{c|}{MIMIC-CXR} & \multicolumn{3}{c|}{OpenI} & \multicolumn{3}{c|}{SXY} \\
\cline{2-10}
 & R-1 & R-2 & R-L & R-1 & R-2 & R-L & R-1 & R-2 & R-L \\
\hline
Gemini-1.0-Pro-Vision & \textbf{0.2814} & \textbf{0.1334} & \textbf{0.2259} & 0.1654 & \textbf{0.0663} & 0.1425 & 0.0103 & 0.0000 & 0.0087 \\
Gemini-1.5-Pro & 0.1759 & 0.0624 & 0.1265 & 0.1364 & 0.0505 & 0.1160 & 0.0344 & 0.0124 & 0.0336 \\
Gemini-1.5-Flash & 0.1973 & 0.0852 & 0.1488 & 0.1018 & 0.0322 & 0.0811 & 0.0269 & 0.0020 & 0.0250 \\
GPT-3.5-Turbo & 0.2406 & 0.1152 & 0.1914 & 0.1529 & 0.0554 & 0.1291 & 0.1010 & 0.0305 & 0.0957 \\
GPT-4-Turbo & 0.1235 & 0.0478 & 0.0925 & 0.0800 & 0.0260 & 0.0647 & 0.2298 & 0.0837 & 0.2173 \\
GPT-4o & 0.2275 & 0.0997 & 0.1752 & \textbf{0.1713} & 0.0622 & \textbf{0.1466} & \textbf{0.2805} & 0.0746 & \textbf{0.2635} \\
Yi-Large & 0.0850 & 0.0346 & 0.0659 & 0.0501 & 0.0153 & 0.0411 & 0.2321 & \textbf{0.0968} & 0.2200 \\
Yi-Large-Turbo & 0.1699 & 0.0728 & 0.1303 & 0.0986 & 0.0329 & 0.0816 & 0.1272 & 0.0518 & 0.1185 \\
Claude-3-Opus & 0.1434 & 0.0608 & 0.1080 & 0.0840 & 0.0276 & 0.0704 & 0.0325 & 0.0070 & 0.0299 \\
Llama 3 & 0.1884 & 0.0791 & 0.1429 & 0.1174 & 0.0371 & 0.0971 & 0.0786 & 0.0173 & 0.0740 \\ 
\hline
\end{tabular}%
}
\end{table}

\subsection{Model Generation Time}

\begin{figure}[!h]
    \centering    \includegraphics[width=0.925\textwidth,height=0.59\textwidth,keepaspectratio]{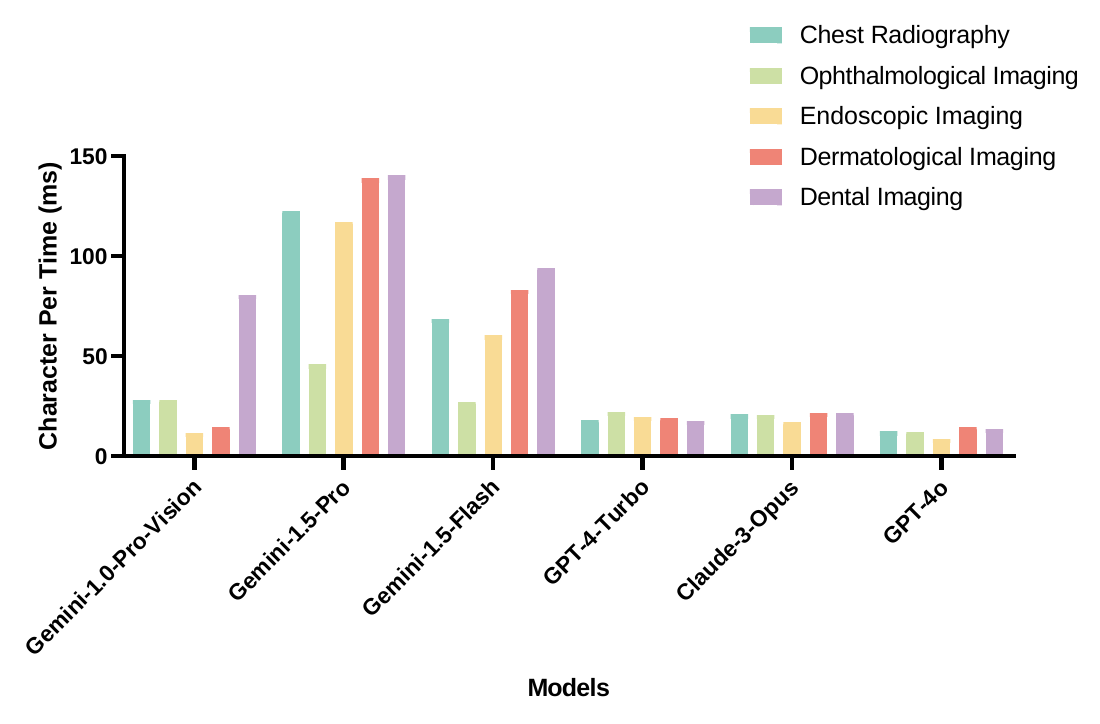}
    \caption{\textbf{Comparison of model generation speed.} All comparative tests were conducted under identical network conditions.}
    \label{fig:time map}
\end{figure}

Given the importance of timeliness in practical applications, our research involved conducting comparative tests on the model's generation speed using the online platform provided by the model within the same network environment. Specifically, we quantified the total number of characters generated by the model and the time taken for this generation process. Subsequently, we calculated the time required for the model to generate each character, as shown from \textbf{Eq. \eqref{eq2}}. As reported in \textbf{Fig. \ref{fig:time map}}, across the five categories of medical image question answering tasks, GPT-4o exhibited the fastest generation speed, except for tasks related to skin image testing. In this particular domain, Gemini-1.0-Pro-Vision demonstrated the fastest performance, with GPT-4o closely following.Gemini-1.5-Pro exhibited the slowest generation speed, with its average time to produce the same number of characters being 9.16 times longer than that of GPT-4o.

% Recognizing the importance of timeliness in practical applications, our research involved comparative tests on the model's generation speed using the online platform within the same network environment. We quantified the total number of characters generated and the time taken for this process, calculating the time required per character, as shown in \textbf{Eq. \eqref{eq2}}. As reported in \textbf{Fig. \ref{fig:time map}}, GPT-4o exhibited the fastest generation speed across five medical image question-answering categories, except for skin image testing, where Gemini-1.0-Pro-Vision performed the fastest, with GPT-4o closely following.Gemini-1.5-Pro exhibited the slowest generation speed, with its average time to produce the same number of characters being 9.16 times longer than that of GPT-4o.

\begin{equation}
\label{eq2}
\text{Character Per Time (ms)} = \frac{\text{Total generation time (ms)}}{\text{Number of characters generated}}
\end{equation}

% \begin{figure}[!h]
%     \centering    \includegraphics[width=0.925\textwidth,height=!,keepaspectratio]{figures/Time map-v2.pdf}
%     \caption{\textbf{Comparison of model generation speed.} All comparative tests were conducted under identical network conditions.}
%     \label{fig:time map}
% \end{figure}

% \newpage

\section{Discussion and Conclusion}

In this comprehensive study, we rigorously evaluated the performance of 10 prominent large models in medical image understanding and radiology report generation, including globally leading models such as Gemini-1.5-Pro, GPT-4o, Claude-3-Opus, Yi-Large-Turbo, among others. 

Our assessment benchmarks these models in explaining medical images, summarizing their advantages and disadvantages, and exploring their potential in medical applications. The findings indicate that while current state-of-the-art MLLMs cannot yet be directly applied to the medical field, their robust reasoning abilities and impressive response speed suggest significant potential for improving model generalization in this domain.

Additionally, we benchmarked these models in generating radiological reports to understand their varying capabilities, strengths, and weaknesses. Our results affirm the performance of numerous domestic and international MLLMs, highlighting their untapped potential in healthcare, especially in radiology. These insights indicate a promising development trajectory, with multilingual and diverse MLLMs poised to enhance global healthcare systems. 

Looking ahead, our large-scale research provides a foundation for further exploration, suggesting the potential to extend these MLLMs to different medical specialties and develop multimodal medical MLLMs for comprehensive patient health understanding. However, ethical considerations such as privacy protection, model fairness, and interpretability, along with legal and regulatory frameworks, are essential for safe and ethical MLLM deployment in healthcare. In summary, despite the promise of reducing doctors' workloads and alleviating medical resource constraints, significant enhancements and comprehensive validation are urgently needed before clinical deployment of these MLLMs.

\newpage

\bibliography{LLM_radoncphys}
\bibliographystyle{unsrt}

\newpage
\appendix
\section{Appendix}
\subsection{Chest Radiography}

\begin{figure}[!htbp]
    \centering    \includegraphics[width=0.8\textwidth,height=!,keepaspectratio]{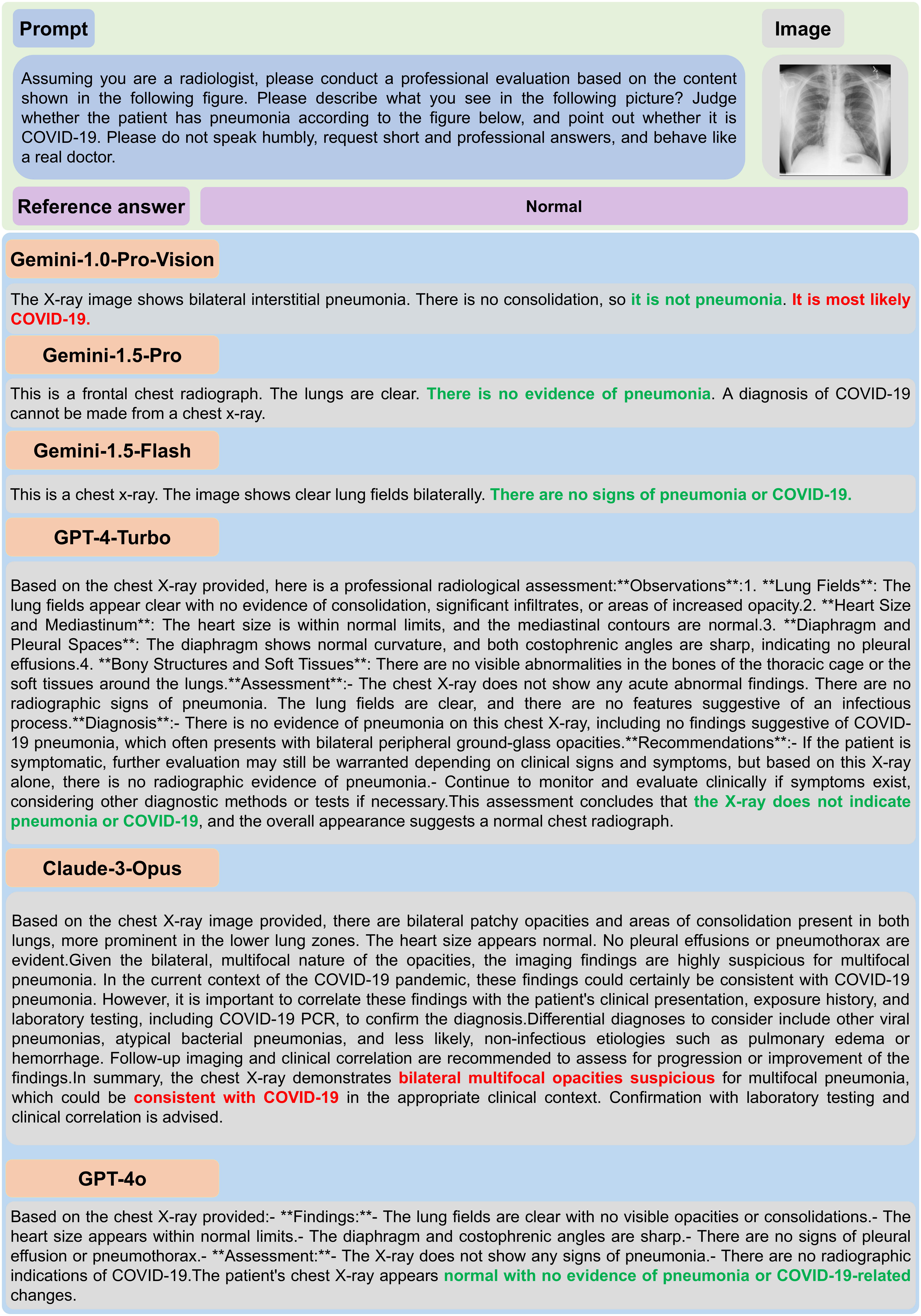}
    \caption{\textbf{Chest: Case 2.} In the context of the three-class classification tasks for pneumonia (normal, pneumonia, COVID-19), the green annotation indicates the correctly identified segments, while the red annotation denotes the incorrectly identified segments.}
    \label{fig:chestXray-case2}
\end{figure}

\begin{figure}[H]
    \centering    \includegraphics[width=0.9\textwidth,height=!,keepaspectratio]{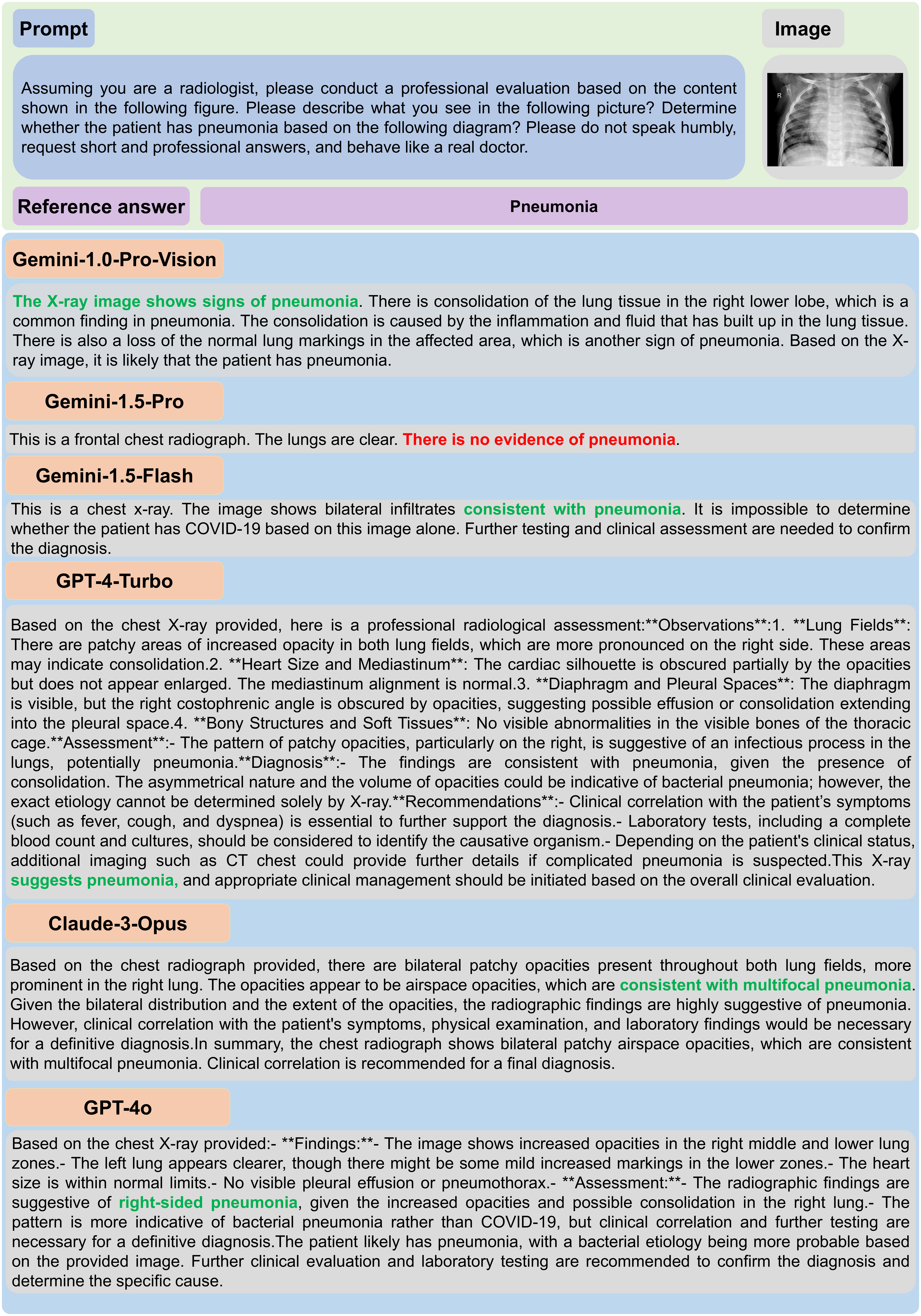}
    \caption{\textbf{Chest: Case 3.} In the context of the binary classification task for pneumonia (normal, pneumonia), green annotations indicate correctly recognized fragments, whereas red annotations indicate incorrectly recognized fragments.}
    \label{fig:chestXray-case3}
\end{figure}

\begin{figure}[H]
    \centering    \includegraphics[width=0.9\textwidth,height=!,keepaspectratio]{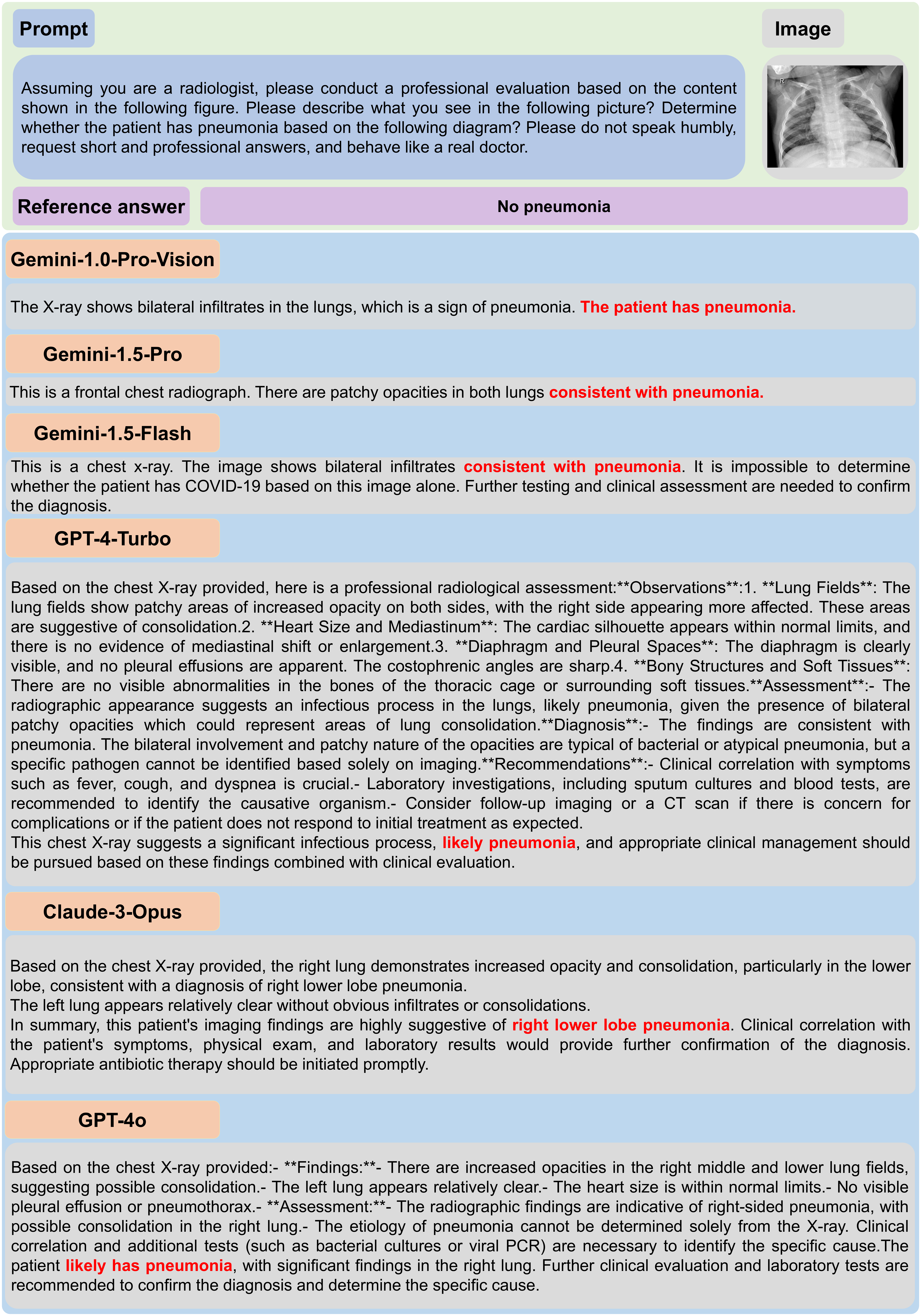}
    \caption{\textbf{Chest: Case 4.} In the context of the binary classification task for pneumonia (normal, pneumonia), green annotations indicate correctly recognized fragments, whereas red annotations indicate incorrectly recognized fragments.}
    \label{fig:chestXray-case4}
\end{figure}

\subsection{Ophthalmological Imaging}

\begin{figure}[!htbp]
    \centering    \includegraphics[width=0.8\textwidth,height=!,keepaspectratio]{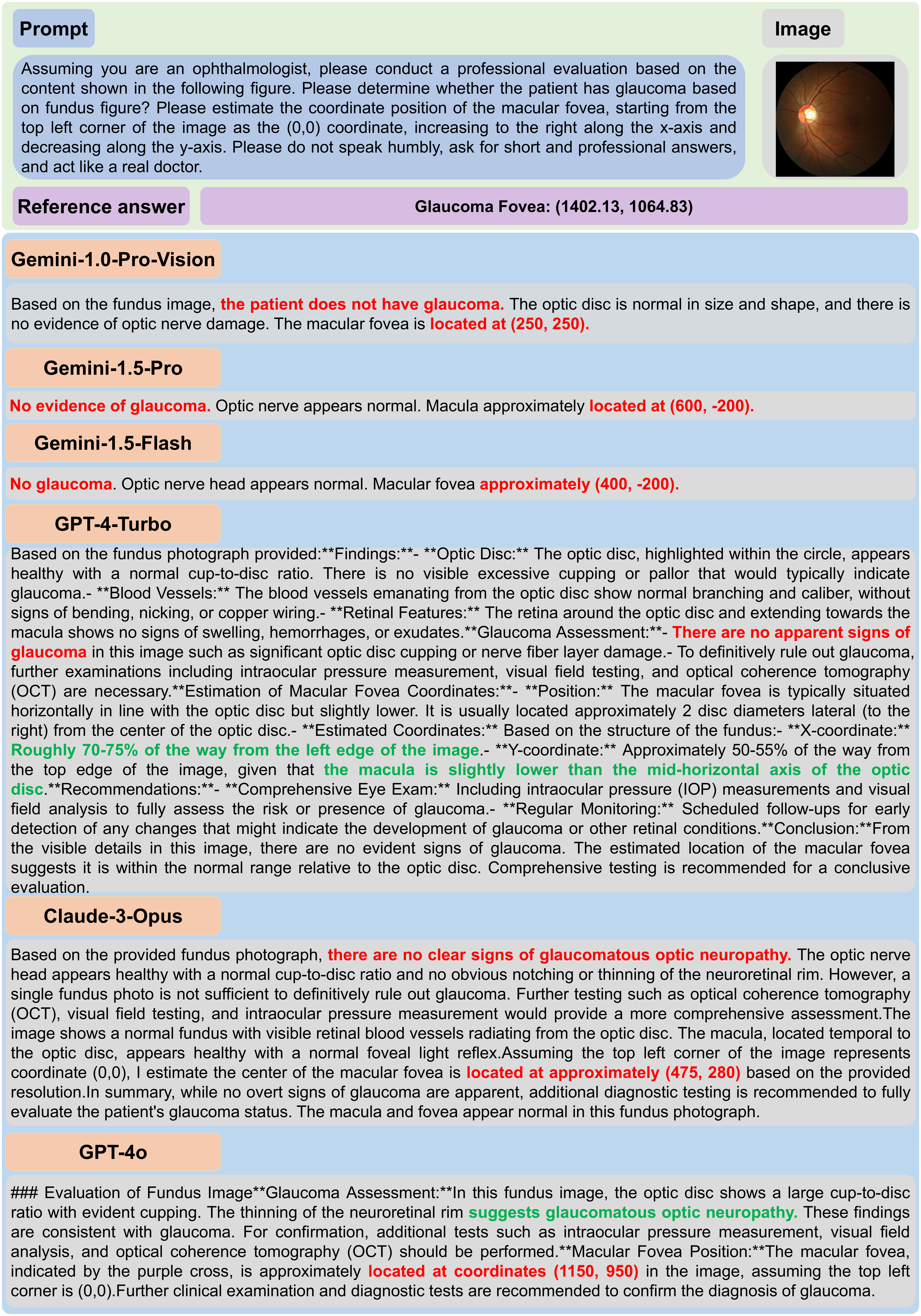}
    \caption{\textbf{Ophthalmological Imaging: Case 2.} Diagnosis of glaucoma and the task of localizing the macular fovea,  the green annotation indicates the correctly identified segments, while the red annotation denotes the incorrectly identified segments.}
    \label{fig:eye-case3}
\end{figure}

\begin{figure}[H]
    \centering    \includegraphics[width=0.9\textwidth,height=!,keepaspectratio]{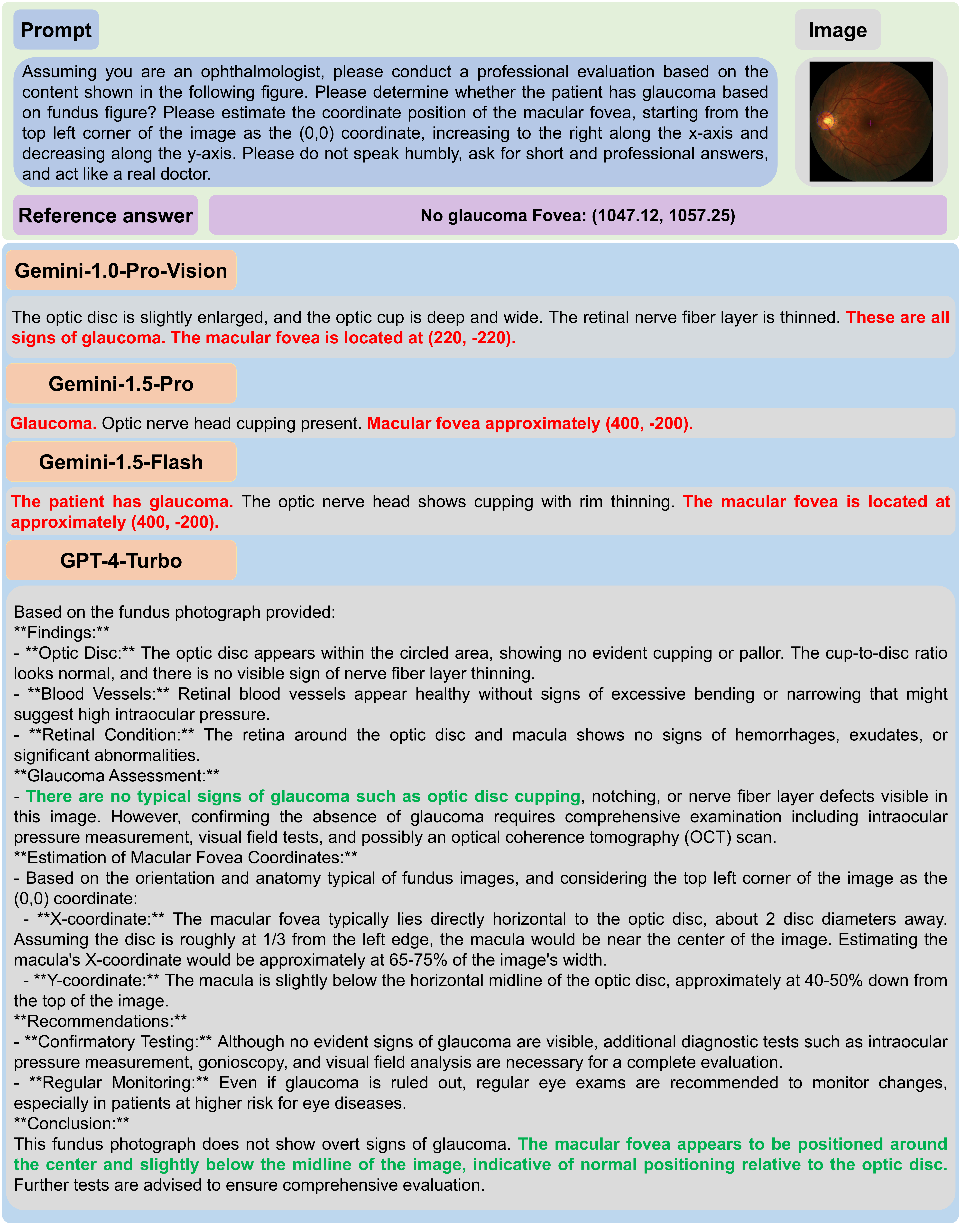}
    \caption{\textbf{Ophthalmological Imaging: Case 3(Part 1).} Diagnosis of glaucoma and the task of localizing the macular fovea,  the green annotation indicates the correctly identified segments, while the red annotation denotes the incorrectly identified segments.}
    \label{fig:eye-case2_1}
\end{figure}

\begin{figure}[H]
    \centering    \includegraphics[width=0.9\textwidth,height=!,keepaspectratio]{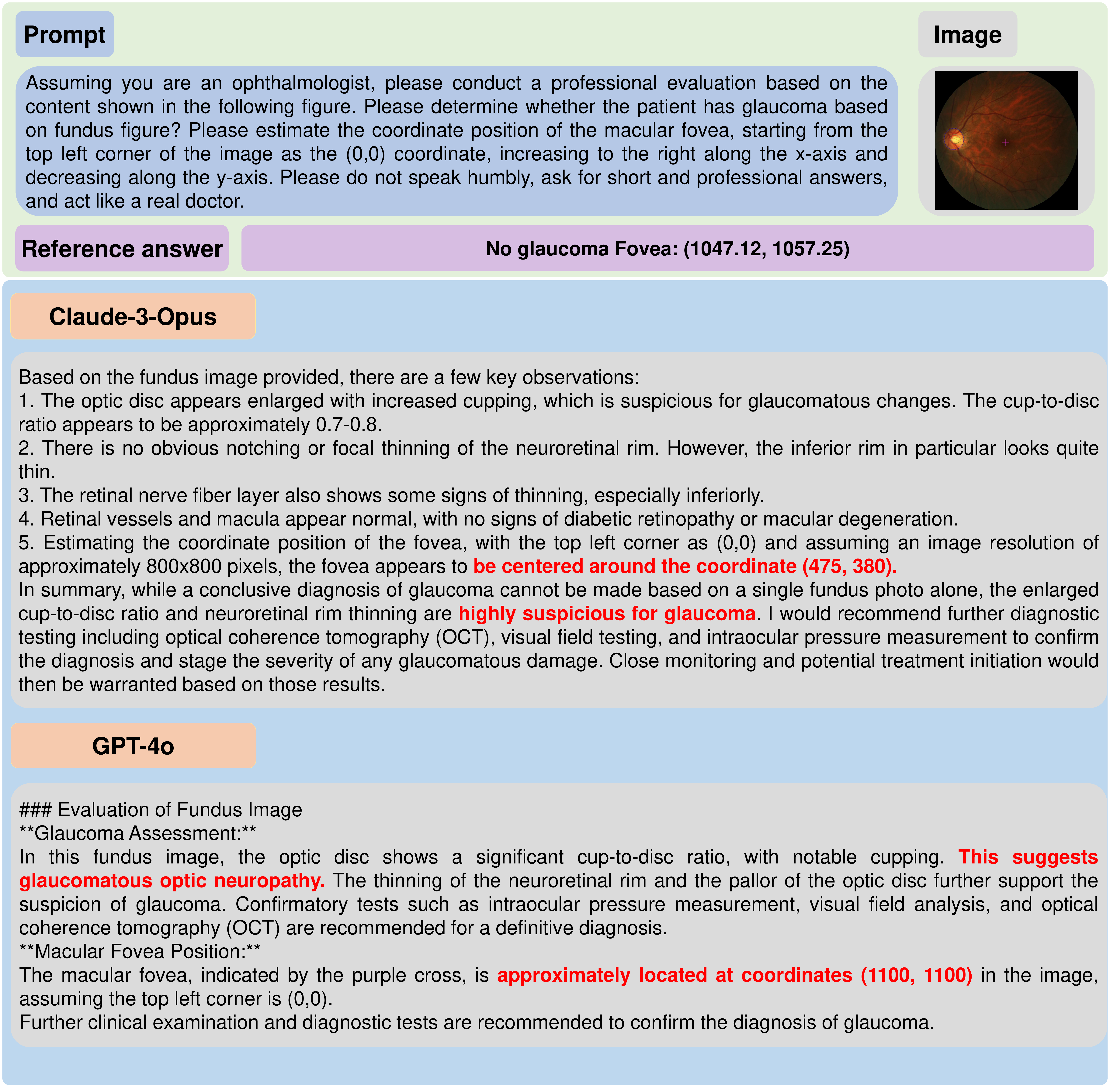}
    \caption{\textbf{Ophthalmological Imaging: Case 3(Part 2).} Diagnosis of glaucoma and the task of localizing the macular fovea,  the green annotation indicates the correctly identified segments, while the red annotation denotes the incorrectly identified segments.}
    \label{fig:eye-case2_2}
\end{figure}

\begin{figure}[H]
    \centering    \includegraphics[width=0.9\textwidth,height=!,keepaspectratio]{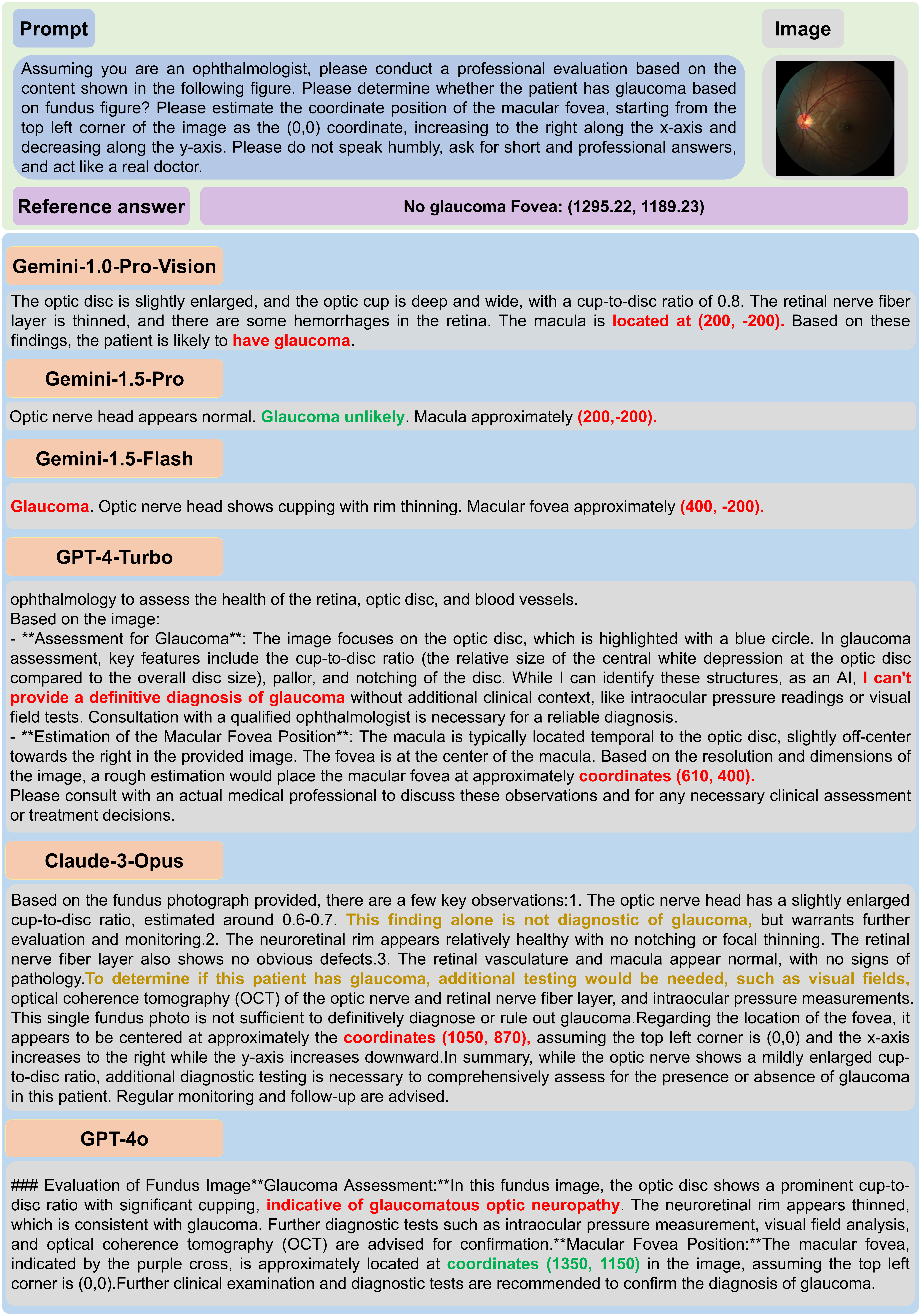}
    \caption{\textbf{Ophthalmological Imaging: Case 4.} Diagnosis of glaucoma and the task of localizing the macular fovea,  the green annotation indicates the correctly identified segments, while the red annotation denotes the incorrectly identified segments.}
    \label{fig:eye-case4}
\end{figure}

\subsection{Endoscopic Imaging}

\begin{figure}[!htbp]
    \centering    \includegraphics[width=0.9\textwidth,height=!,keepaspectratio]{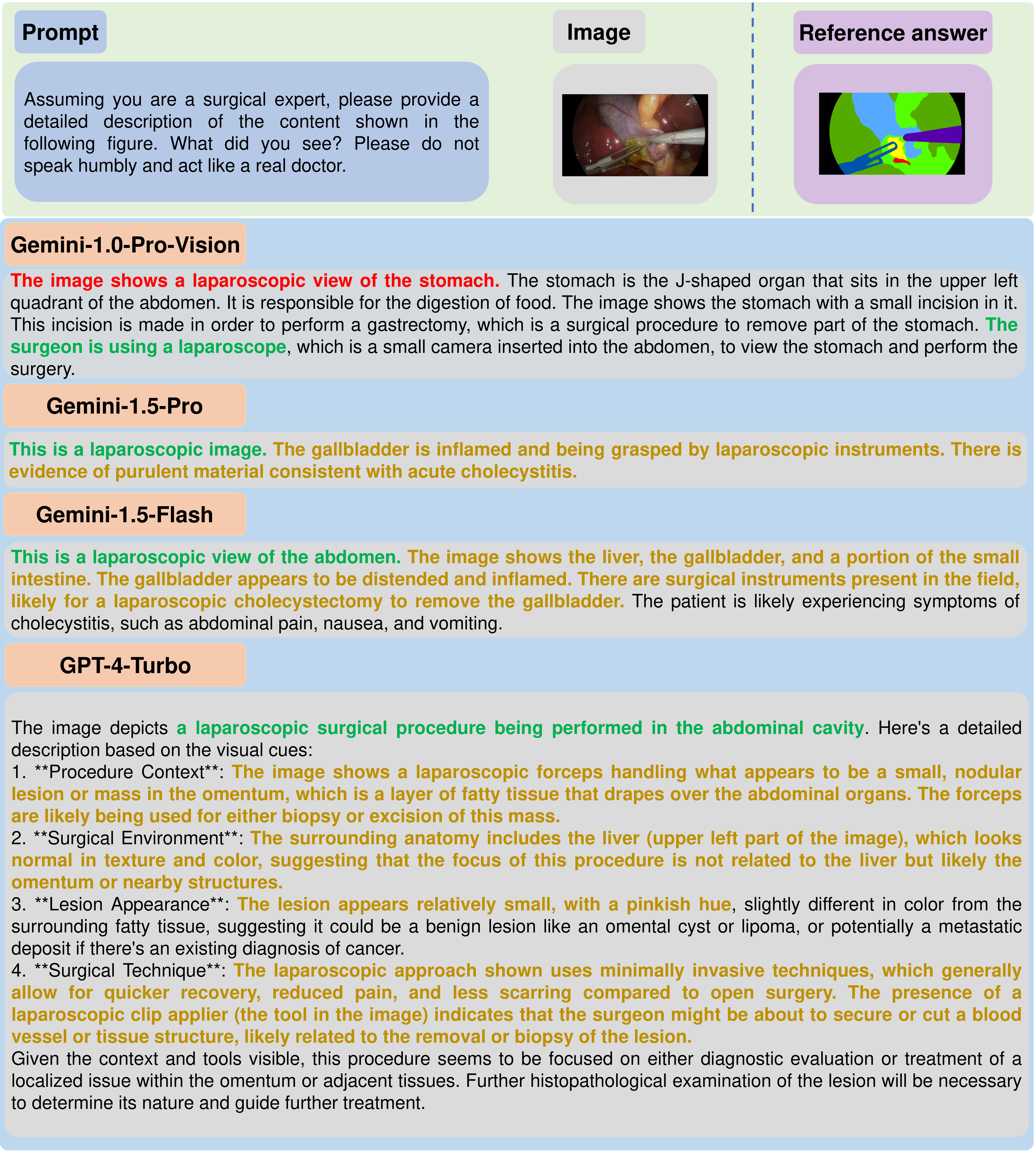}
    \caption{\textbf{Endoscopic: Case 2(Part 1).} The task of abdominal segmentation during the surgical process employs green markings to denote correct answers, red markings to indicate incorrect answers, and yellow markings to highlight noteworthy content.}
    \label{fig:endoscopic-case2_1}
\end{figure}

\begin{figure}[!htbp]
    \centering    \includegraphics[width=0.9\textwidth,height=!,keepaspectratio]{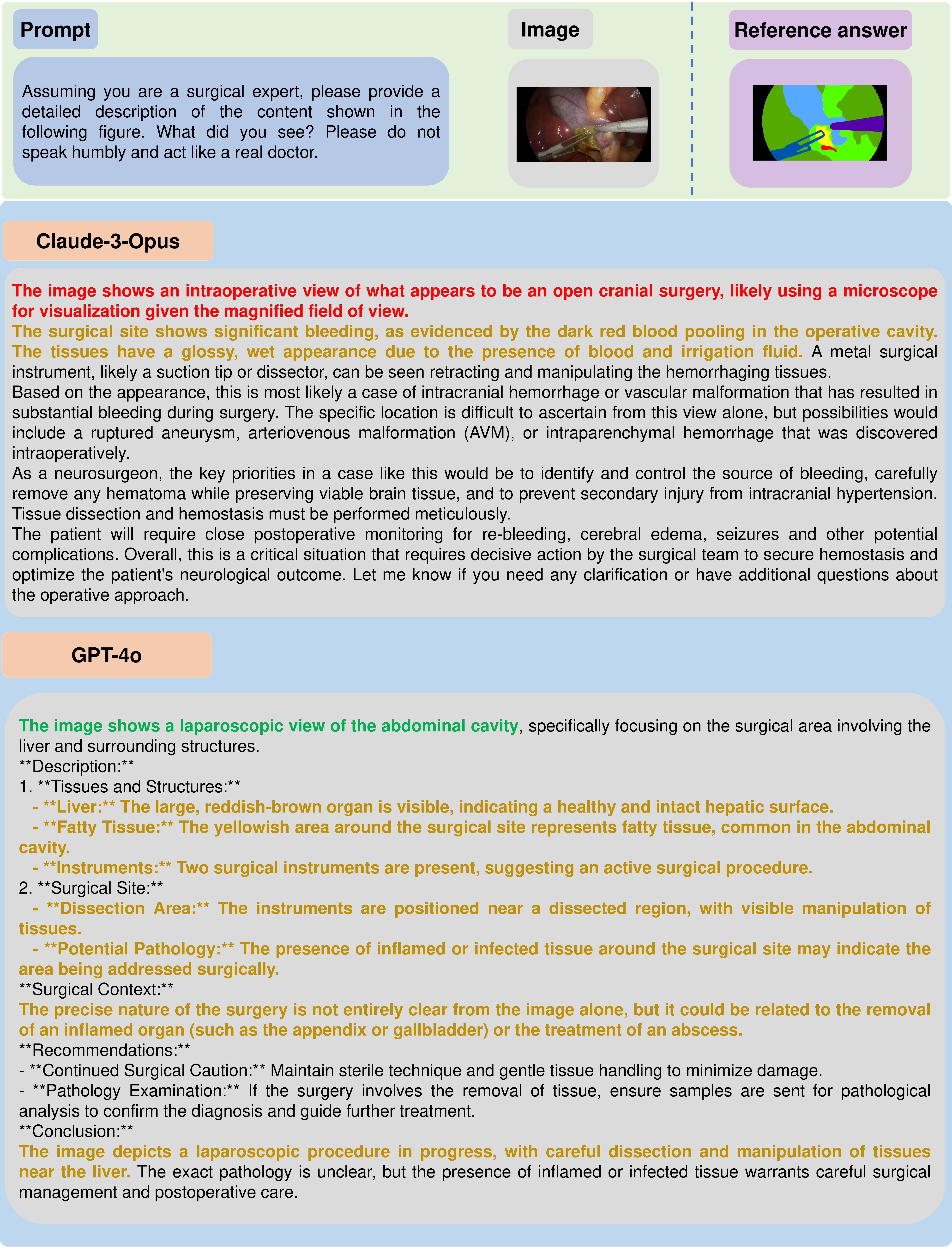}
    \caption{\textbf{Endoscopic: Case 2(Part 2).} The task of abdominal segmentation during the surgical process employs green markings to denote correct answers, red markings to indicate incorrect answers, and yellow markings to highlight noteworthy content.}
    \label{fig:endoscopic-case2_2}
\end{figure}

\begin{figure}[H]
    \centering    \includegraphics[width=0.9\textwidth,height=!,keepaspectratio]{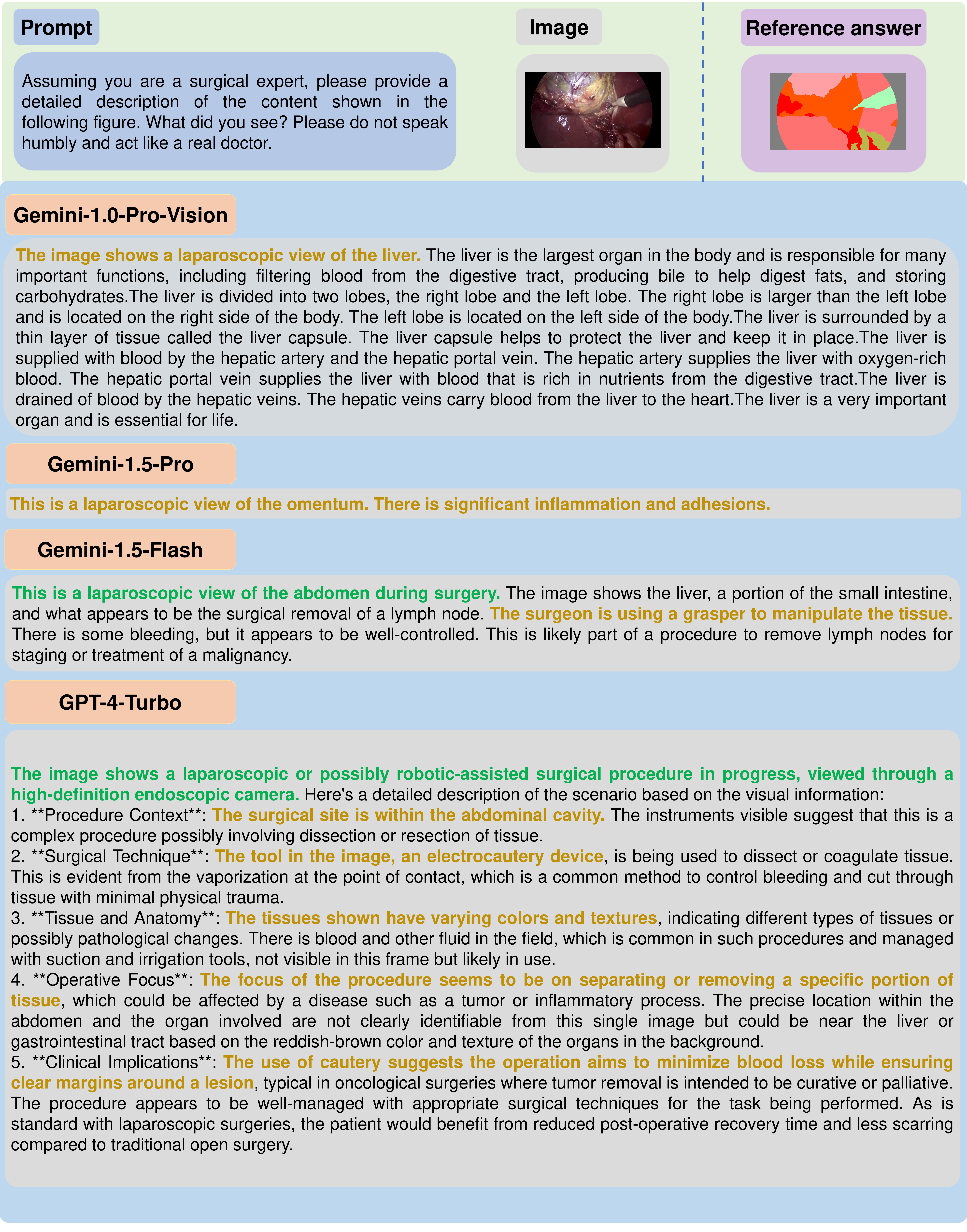}
    \caption{\textbf{Endoscopic: Case 3(Part 1).} The segmentation task during cholecystectomy surgery uses green markings to indicate correct answers, red markings to indicate incorrect answers, and yellow markings to highlight noteworthy content.}
    \label{fig:endoscopic-case3_1}
\end{figure}

\begin{figure}[H]
    \centering    \includegraphics[width=0.9\textwidth,height=!,keepaspectratio]{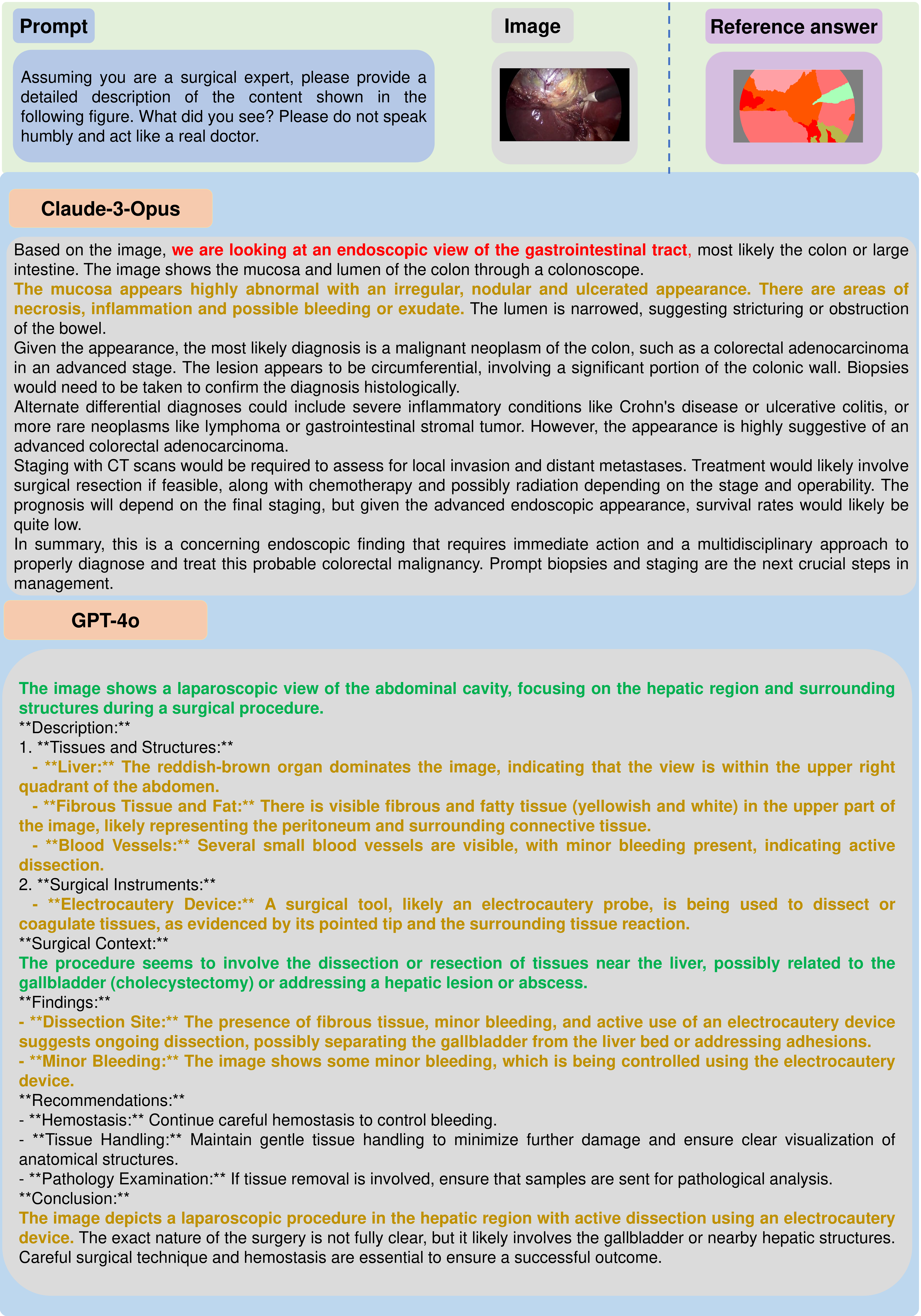}
    \caption{\textbf{Endoscopic: Case 3(Part 2).} The segmentation task during cholecystectomy surgery uses green markings to indicate correct answers, red markings to indicate incorrect answers, and yellow markings to highlight noteworthy content.}
    \label{fig:endoscopic-case3_2}
\end{figure}

\begin{figure}[H]
    \centering    \includegraphics[width=0.9\textwidth,height=!,keepaspectratio]{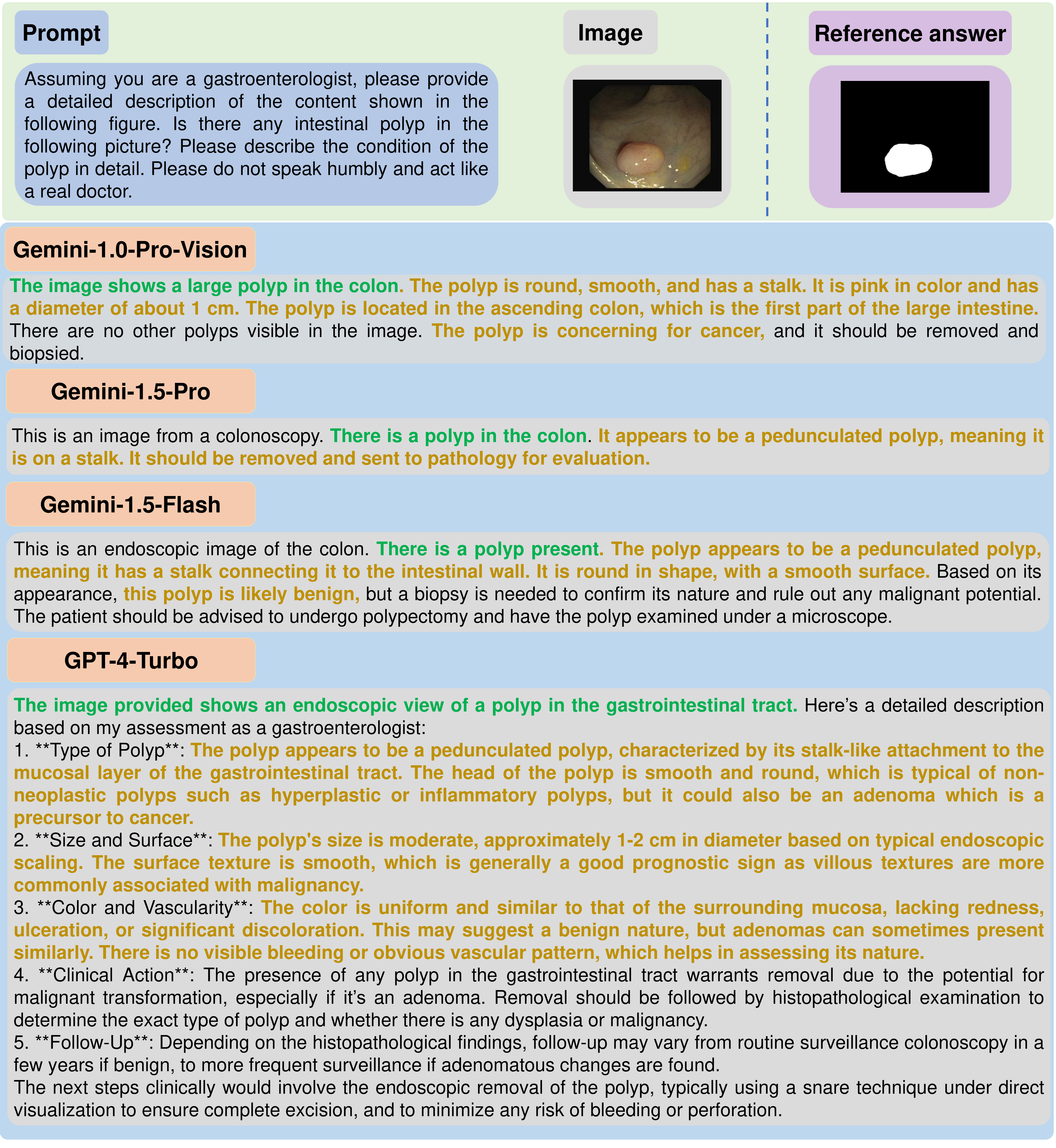}
    \caption{\textbf{Endoscopic: Case 4(Part 1).} The task of colon polyp localization and segmentation uses green markings to represent correct answers, red markings to represent incorrect answers, and yellow markings to highlight noteworthy content.}
    \label{fig:endoscopic-case4_1}
\end{figure}

\begin{figure}[H]
    \centering    \includegraphics[width=0.9\textwidth,height=!,keepaspectratio]{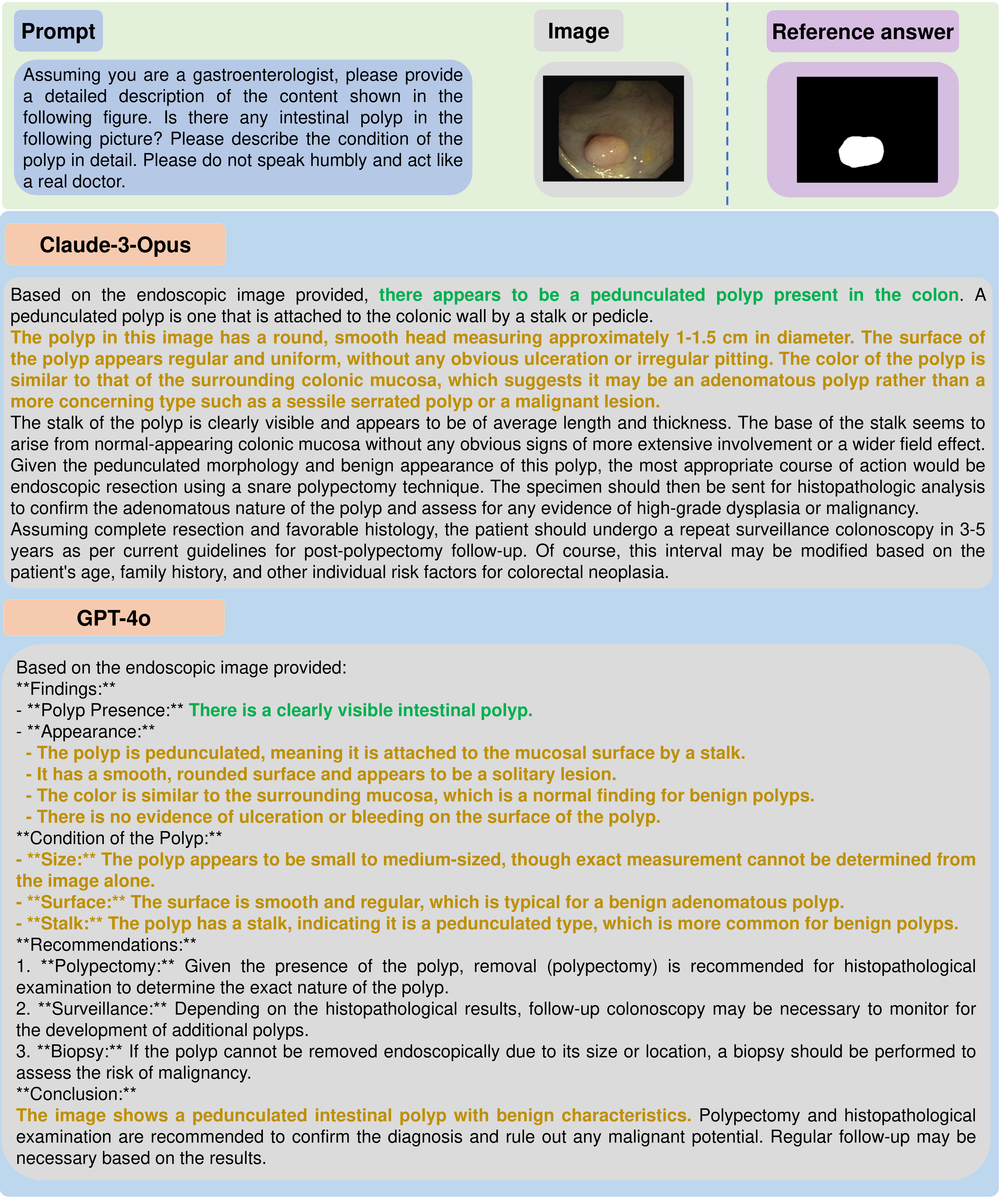}
    \caption{\textbf{Endoscopic: Case 4(Part 2).} The task of colon polyp localization and segmentation uses green markings to represent correct answers, red markings to represent incorrect answers, and yellow markings to highlight noteworthy content.}
    \label{fig:endoscopic-case4_2}
\end{figure}

\subsection{Dermatological Imaging}

\begin{figure}[H]
    \centering    \includegraphics[width=0.85\textwidth,height=!,keepaspectratio]{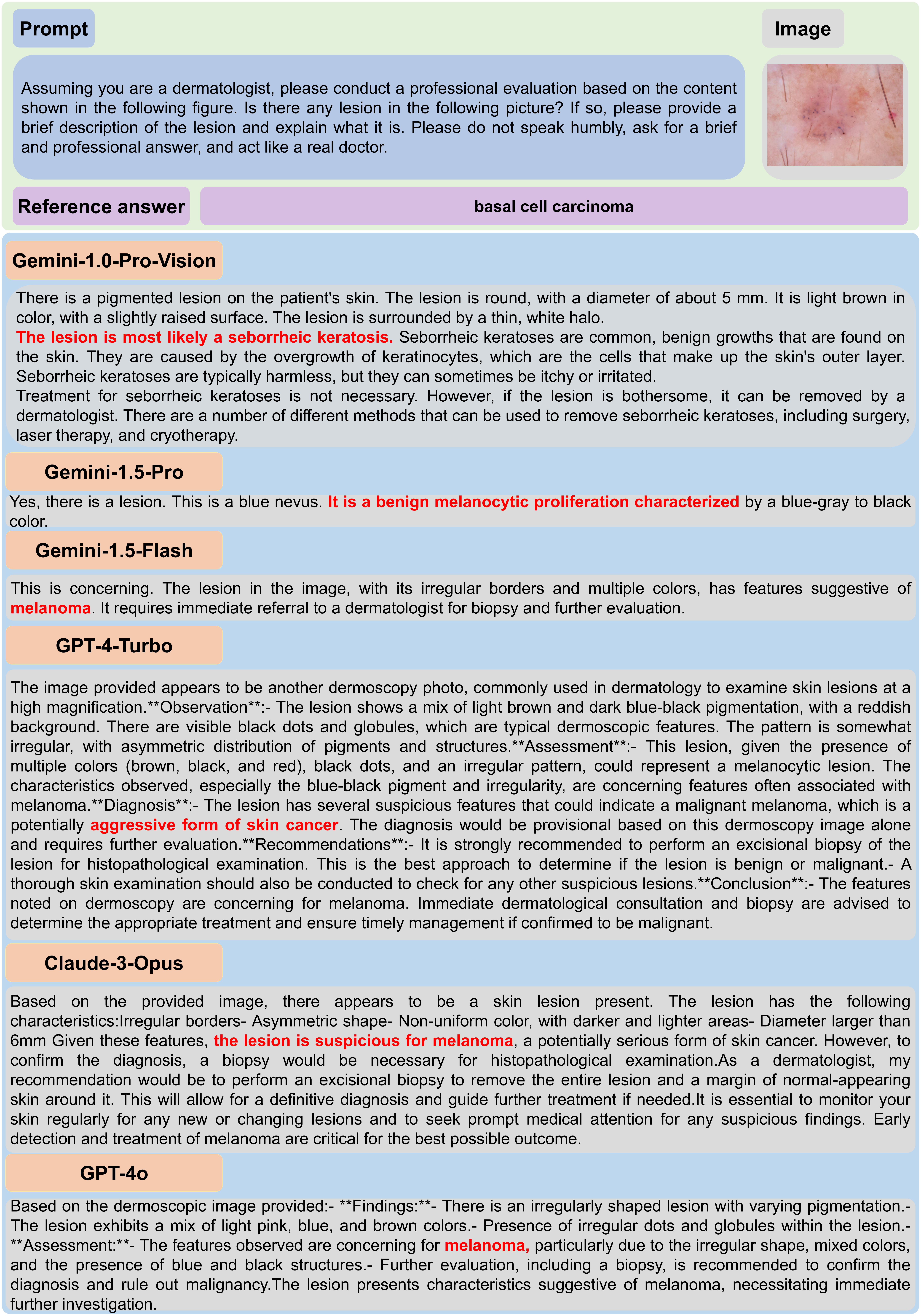}
    \caption{\textbf{Skin: Case 2.} The nine classification tasks for skin diseases utilize green markings to denote correct answers and red markings to indicate incorrect answers.}
    \label{fig:skin-case2}
\end{figure}

\begin{figure}[H]
    \centering    \includegraphics[width=0.85\textwidth,height=!,keepaspectratio]{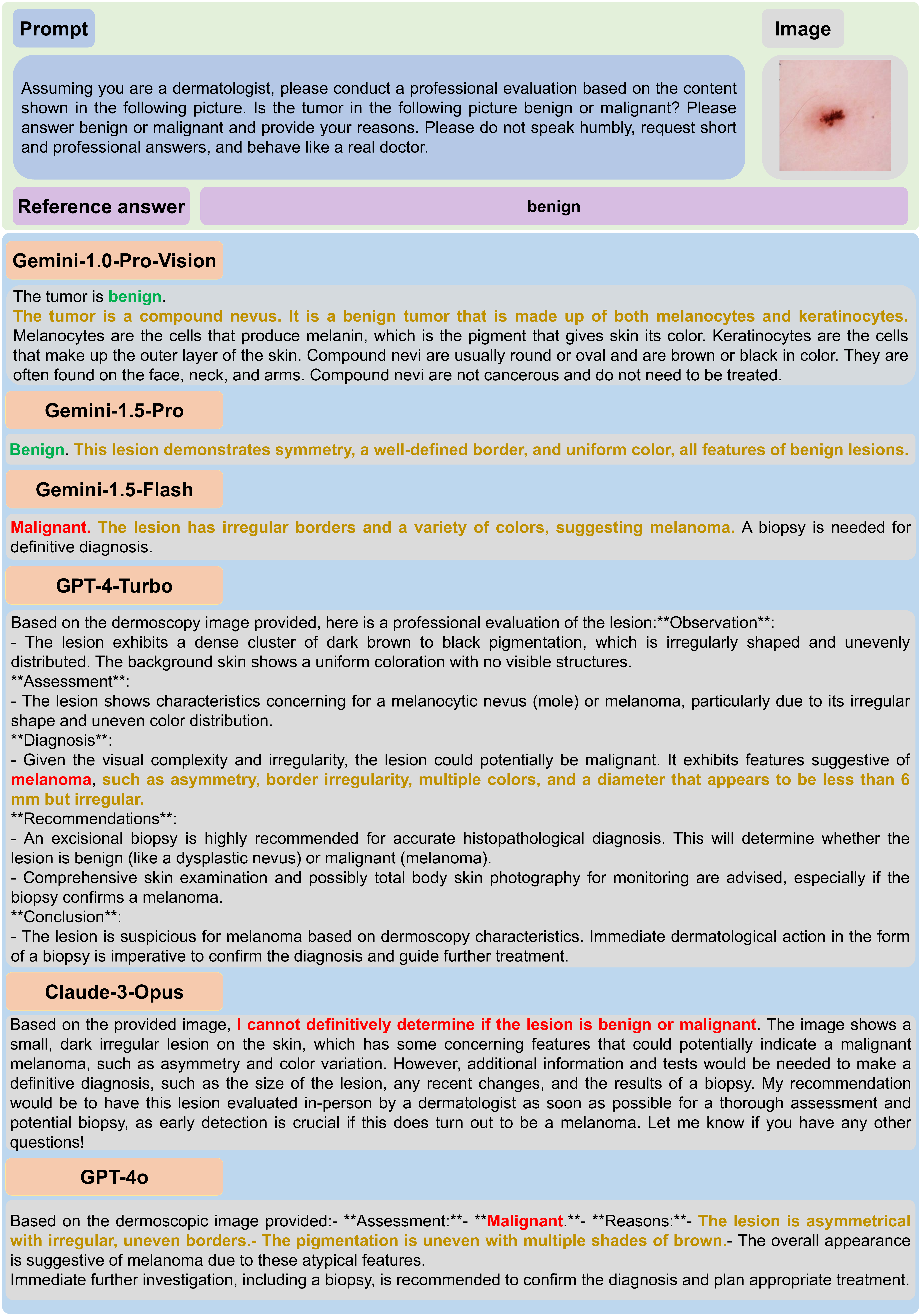}
    \caption{\textbf{Skin: Case 3.} The benign and malignant classification task of skin nevi employs green markers to denote correct answers, red markers to denote incorrect answers, and yellow markers to highlight noteworthy content.}
    \label{fig:skin-case3}
\end{figure}

\begin{figure}[H]
    \centering    \includegraphics[width=0.85\textwidth,height=!,keepaspectratio]{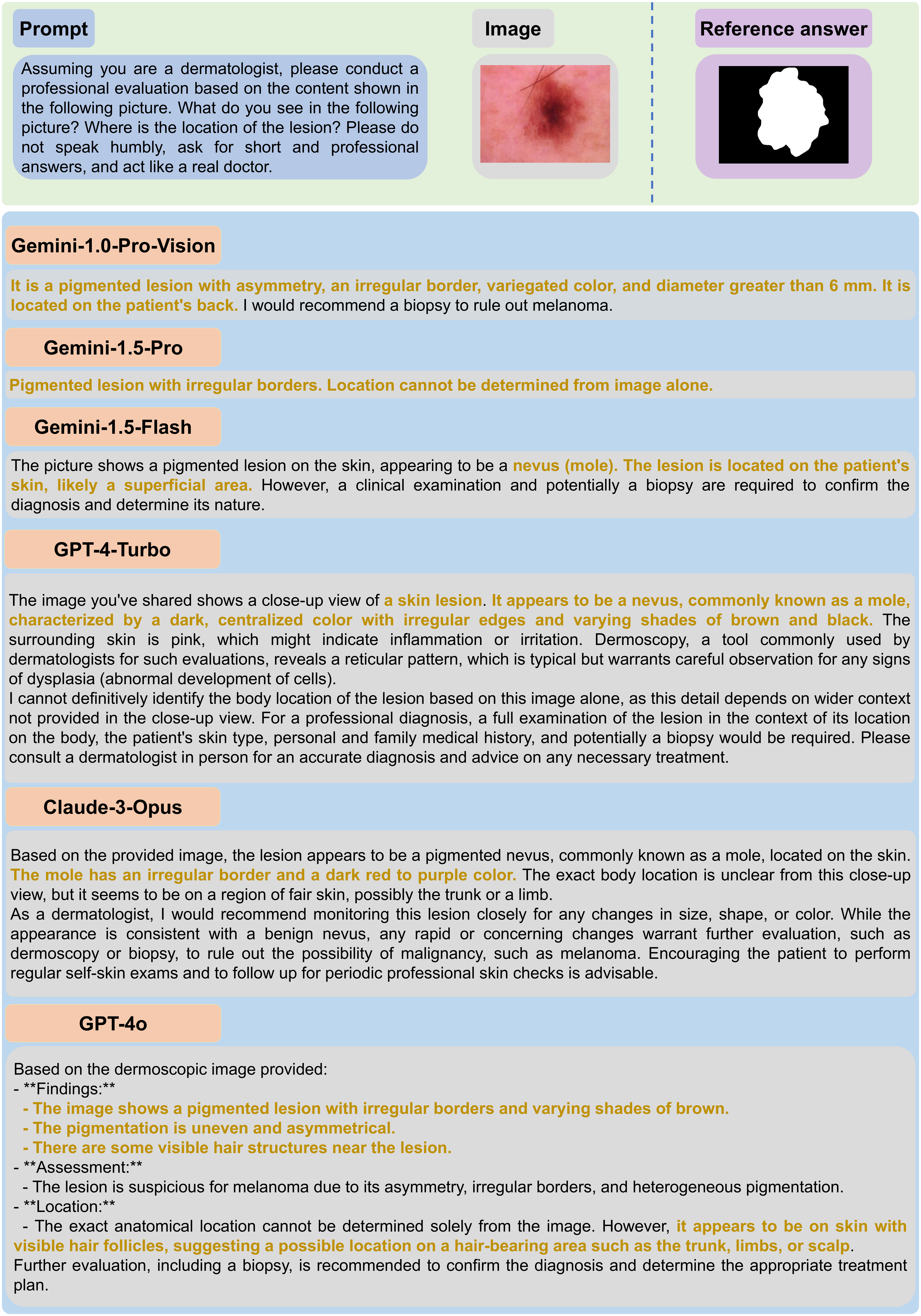}
    \caption{\textbf{Skin: Case 4.} The seven classification recognition and detection task of skin diseases utilizes yellow markers to emphasize noteworthy content.}
    \label{fig:skin-case4}
\end{figure}

\subsection{Dental Imaging}

\begin{figure}[H]
    \centering    \includegraphics[width=0.85\textwidth,height=!,keepaspectratio]{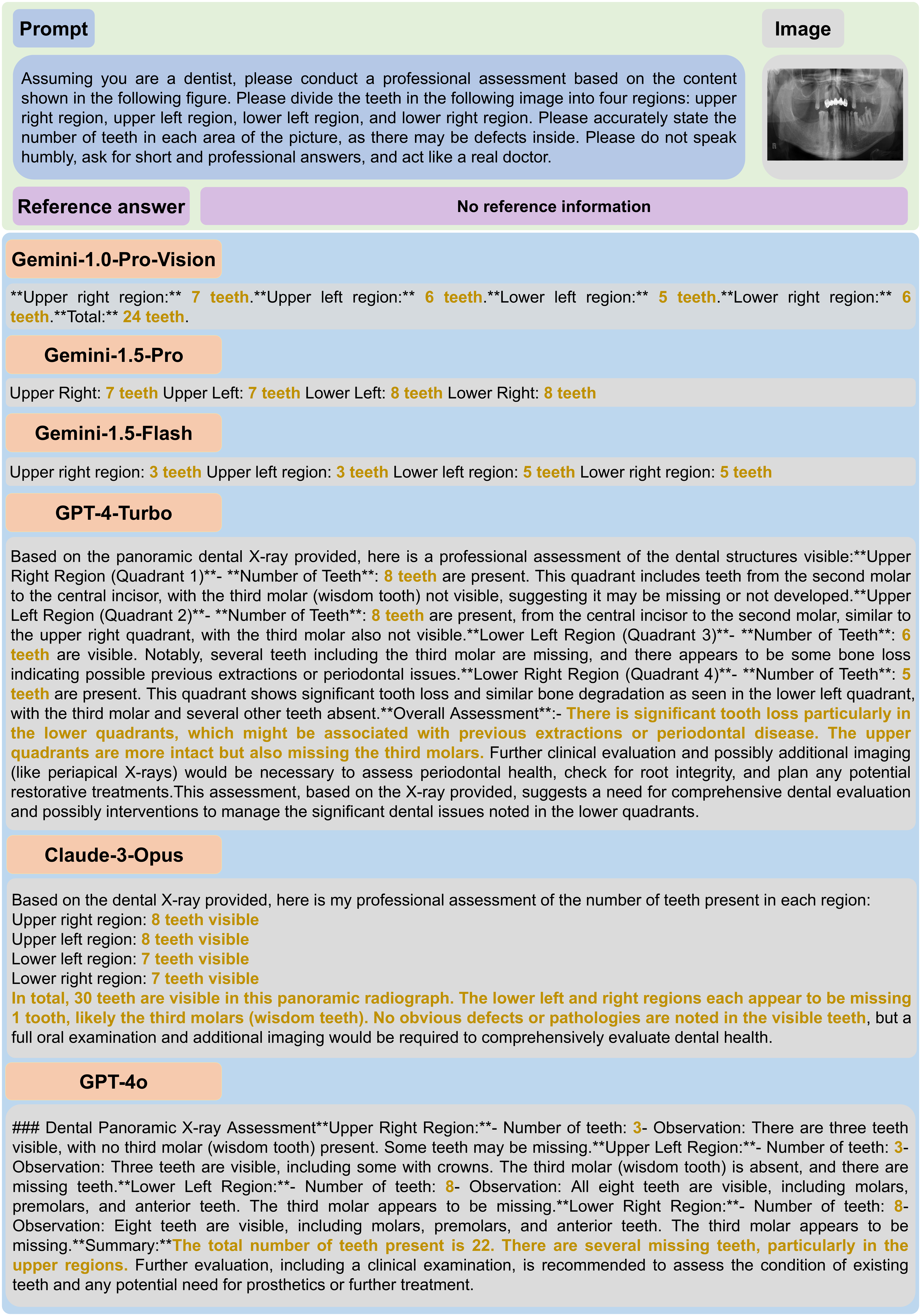}
    \caption{\textbf{Dental Image: Case 2.} The task involves quantifying the number of teeth and identifying anomalies in dental X-rays, with critical information highlighted in yellow.}
    \label{fig:dentalXray-case2}
\end{figure}

\begin{figure}[H]
    \centering    \includegraphics[width=0.9\textwidth,height=!,keepaspectratio]{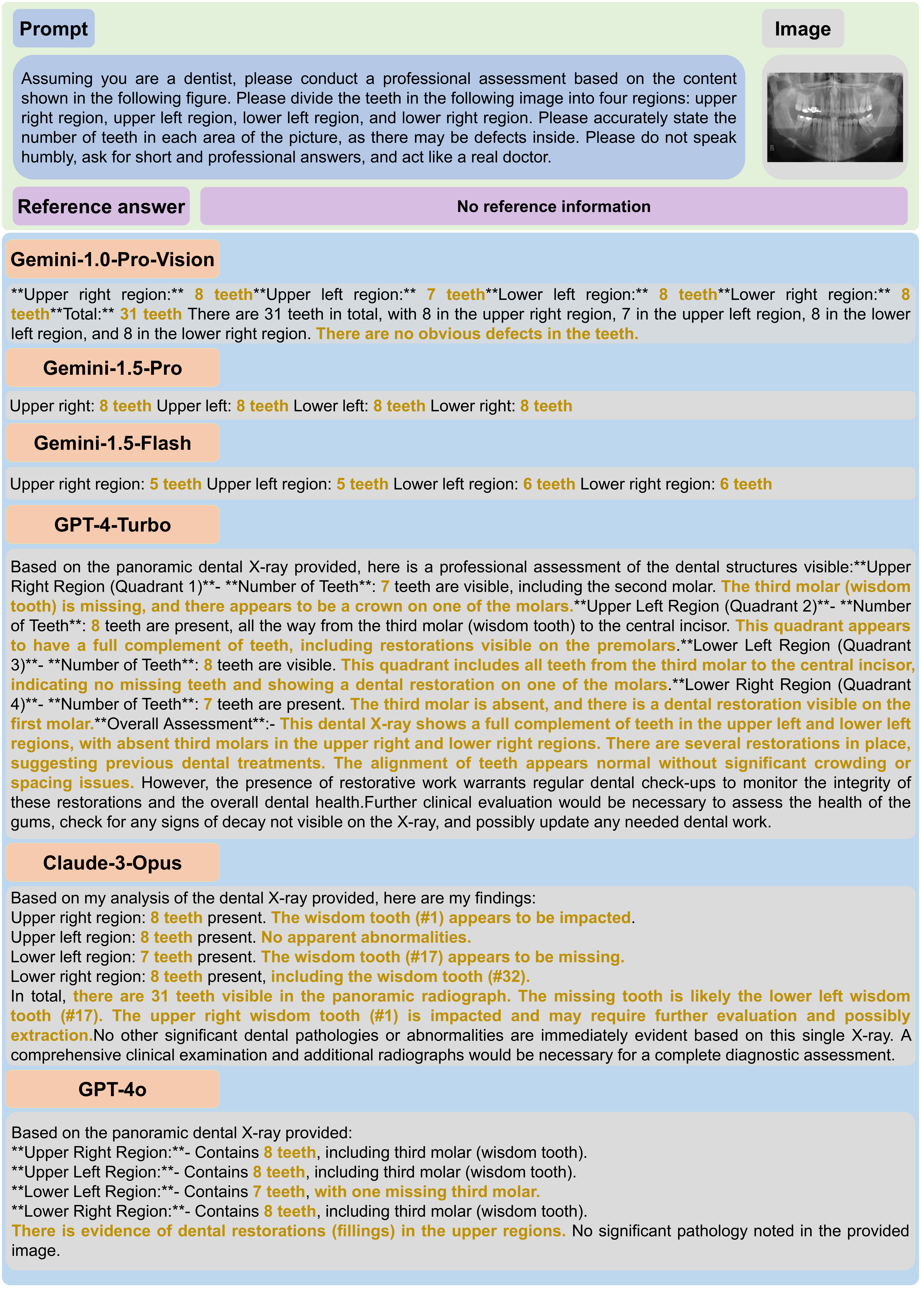}
    \caption{\textbf{Dental Image: Case 3.} The task involves quantifying the number of teeth and identifying anomalies in dental X-rays, with critical information highlighted in yellow.}
    \label{fig:dentalXray-case3}
\end{figure}

\begin{figure}[H]
    \centering    \includegraphics[width=0.9\textwidth,height=!,keepaspectratio]{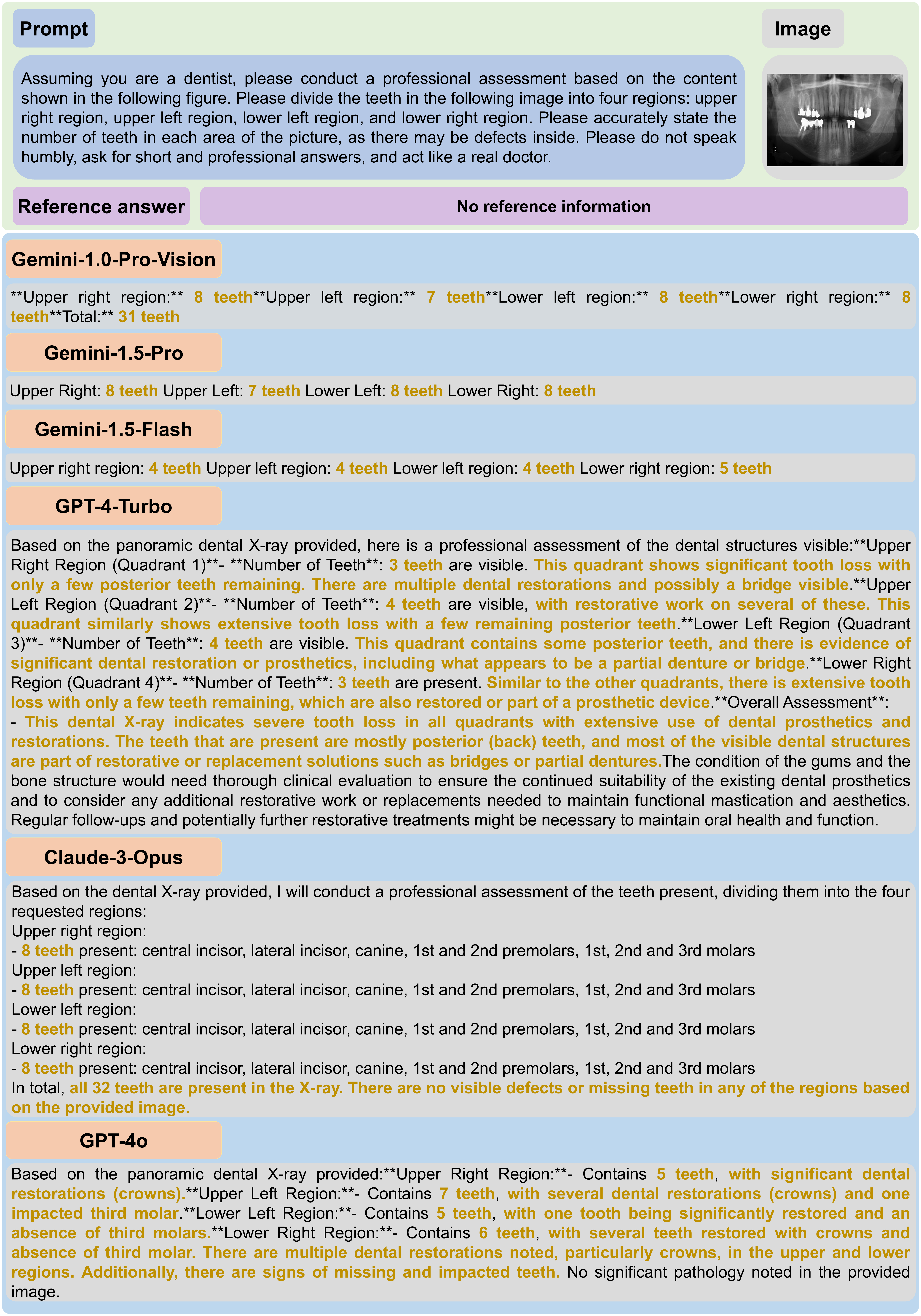}
    \caption{\textbf{Dental Image: Case 4.} The task involves quantifying the number of teeth and identifying anomalies in dental X-rays, with critical information highlighted in yellow.}
    \label{fig:dentalXray-case4}
\end{figure}

\end{document}